\documentclass[12pt]{article}

\usepackage{hyperref}
\usepackage{mathrsfs}
\PassOptionsToPackage{unicode}{hyperref}
\PassOptionsToPackage{naturalnames}{hyperref}

\usepackage{enumitem}
\usepackage{graphicx}
\usepackage{color}
\usepackage{soul}
\usepackage{comment}
\usepackage{epstopdf}
\usepackage{float}
\usepackage{feynmp}
\usepackage{slashed}
\usepackage{tikz}   
\usepackage{tikz-feynman}
\usepackage{soul}
\usepackage{cite}
\usepackage{subfig}
\usepackage{amstext,amssymb}
\usepackage{graphicx}
\usepackage[section]{placeins}
\usepackage{xspace}
\usepackage{color}
\usepackage{units}
\usepackage{array}
\usepackage[section]{placeins}

\usepackage{amsmath,bm}
\usepackage{amssymb}
\usepackage{gensymb}
\usepackage{soul}
\usepackage{textcomp}
\usepackage{slashed} % feynman 
\usepackage{multirow}
\newcommand{\be}{\begin{equation}}
\newcommand{\ee}{\end{equation}}
\newcommand{\bea}{\begin{eqnarray}}
\newcommand{\eea}{\end{eqnarray}}
\newcommand{\nn}{\nonumber}
\usepackage{epstopdf}
\usepackage{float}
\usepackage{multirow}

\def\s1{\hat s}

\usepackage{slashed}
\global\long\def\d{\partial}

\newcommand{\PreserveBackslash}[1]{\let\temp=\\#1\let\\=\temp}
\newcolumntype{C}[1]{>{\PreserveBackslash\centering}p{#1}}
\newcolumntype{R}[1]{>{\PreserveBackslash\raggedleft}p{#1}}
\newcolumntype{L}[1]{>{\PreserveBackslash\raggedright}p{#1}}

\usepackage{slashed}

\def\thefootnote{\fnsymbol{footnote}}

\usepackage{hyperref}
\usepackage{slashed}

\usepackage{lscape}%
\usepackage{array}
\usepackage{booktabs}%

\newcommand{\nua}[1]{\ensuremath{\rlap{\kern-2.5pt\ensuremath{\overset{\scriptscriptstyle(-)}{\phantom{\nu}}}}{\ensuremath{{\nu}_{#1}}}}\xspace}

\begin{document}

\vspace{0.2cm}

\begin{center}
{\Large\bf Exploring neutrino masses, $(g-2)_{\mu, e}$ in type I+II seesaw in ${L^{}_e - L^{}_{\alpha}}$ gauge extended model}
\end{center}

\vspace{0.1cm}

\begin{center}
 {\bf Papia Panda$^1$}~\footnote{E-mail: ppapia93@gmail.com},  {\bf Priya Mishra$^1$}~\footnote{E-mail: mishpriya99@gmail.com}, 
 {\bf Mitesh Kumar Behera$^2$}~\footnote{E-mail: miteshbehera1304@gmail.com},
 {\bf Shivaramakrishna Singirala$^{1}$}~\footnote{E-mail: krishnas542@gmail.com}, {\bf Rukmani Mohanta$^1$}~\footnote{E-mail: rmsp@uohyd.ac.in}
\\
\vspace{0.2cm}
{\it $^1$School of Physics, 
University of Hyderabad, Hyderabad, India-500046\\}
{\it $^2$High Energy Physics Research Unit, Department of Physics, Faculty of Science, Chulalongkorn University, Bangkok 10330, Thailand\\}
\end{center}

\vspace{0.05cm}

\begin{abstract}
    This paper aims to explore the implications of  $U(1)_{L_e-L_{\alpha}}$ gauge symmetries, where $\alpha=\tau, \mu$, in the neutrino sector through the type-(I+II) seesaw mechanisms. To achieve such a hybrid framework, we include a scalar triplet and three right-handed neutrinos. The model can successfully account for the active neutrino masses, mixing angles, mass squared differences, and the CP-violating phase within the $3 \sigma$ bounds of NuFit v5.2 neutrino oscillation data. The presence of new gauge boson  at the MeV scale provides an explanation for the muon and electron $(g-2)$ within the confines of their experimental limits. Furthermore, we scrutinize the proposed models in the context of upcoming long-baseline neutrino experiments such as DUNE, P2SO, T2HK, and T2HKK. 
    The findings reveal that P2SO and T2HK have the ability to probe both the models in their $5 \sigma$ allowed oscillation parameter region, whereas DUNE and T2HKK can conclusively test only model with $U(1)_{L_e-L_\mu}$- symmetry within their $5 \sigma$ parameter space if the true values of the oscillation parameters remain consistent with NuFit v5.2.
\end{abstract}

\def\thefootnote{\arabic{footnote}}
\setcounter{footnote}{0}

\newpage

\section{Introduction}
\label{sec:intro}
The success of the Standard Model (SM) in explaining the observed phenomena in particle physics is indubitable; yet, exceptions like dark matter, neutrino masses and mixing, baryon asymmetry of the Universe, etc., cannot be realized within it and are beyond its realm. Among these exceptions, it is necessary to move beyond the standard model (BSM) to explain neutrino masses and mixing. In this context, type-I seesaw \cite{minkowski1977mu, Yanagida:1979as, Gell-Mann:1979vob, Brdar:2019iem, Miranda:2016ptb, Branco:2020yvs,Ramond:1979py} being the simplest and the most elegant one, other modified seesaw mechanisms such as type-II {\cite{Schechter:1980gr, 
 Mohapatra:1980yp, Cheng:1980qt, Ma:1998dx, Antusch:2004xy, Gu:2006wj,Arhrib:2011uy,
 Freitas:2014fda}} via scalar triplet, type-III {\cite{Liao:2009nq, Ma:1998dn, Foot:1988aq, Chen:2011de} }via fermion triplet, linear seesaw {\cite{Malinsky:2005bi, Ma:2009du}}, inverse seesaw {\cite{Bazzocchi:2009kc, Ma:2009gu}}, etc., are being explored phenomenologically in myriad literature. Despite the remarkable precision achieved in determining the masses of charged leptons, our understanding of neutrino masses remains elusive and lacks comparable accuracy \cite{Zyla:2020zbs}. In this work,  we incorporate the type-II seesaw in addition to the type-I by including an additional  $SU(2)_L$ triplet scalar field $\Delta$ to explain neutrino phenomenology.

Even though both the type-I and type-II seesaw mechanisms appear appealing for explaining the neutrino oscillation data, they need more experimental testability as they require new particles with very heavy masses which are beyond the reach of present and future experiments. This motivates us to propose a simplistic yet feasible model to concurrently account for the neutrino masses, electron and muon $(g-2)$. For this, we consider a hybrid seesaw, i.e., type-(I+II),  in a $U(1)^{}_{L_e - L_{\alpha}}$ gauge extension model where $\alpha=\tau, \mu$ and with the inclusion of three right-handed (RH) neutrinos and a scalar triplet. In this work, we focus on two frameworks: $U(1)_{L_e-L_{\tau}}$, named model A, and $U(1)_{L_e-L_{\mu}}$ named model B. Of these, model A can explain neutrino phenomenology and electron $(g-2)$, while model B can  explain muon, electron  $(g-2)$, and neutrino phenomenology. These kind of models have previously been explored in the literature in the context of
ANITA events \cite{Esmaili:2019pcy}, cosmic ray $e^+e^-$ excess \cite{Duan:2017qwj,Cao:2017sju} and flavor-dependent long-range forces from solar and atmospheric neutrinos \cite{Joshipura:2003jh,Bandyopadhyay:2006uh,Bustamante:2018mzu,KumarPoddar:2020kdz}.

Testing the models in the framework of long-baseline neutrino experiments is crucial, hence, we consider the four most promising upcoming long-baseline neutrino experiments for probing our models: DUNE, P2SO, T2HK, and T2HKK. Considering the predicted values of the oscillation parameters from both models, we delineate the variation of $\delta_{CP}^{\rm true}$ with $\theta_{23}^{\rm true}$ and project it into the parameter space of these upcoming experiments.

The structure of this paper is as follows: in Section \ref{sec:framework}, we present the model framework involving type-(I+II) seesaw mechanism with $U(1)^{}_{L^{}_e - L^{}_{\alpha}}$ gauge symmetry; in Section \ref{subsec:model A}, we specifically discuss the $U(1)^{}_{L^{}_e - L^{}_\tau}$ framework (model A) and its implications for the mass matrices (i.e., charged leptons and neutral leptons); in Section \ref{sec:model B}, we demonstrate model B, achieved by implementing $U(1)^{}_{L^{}_e - L^{}_\mu}$ symmetry; proceeding further, in Section \ref{sec:numerical}, a numerical study between the observables of the neutrino sector and the model input parameters is established for both models; 
Section \ref{test-model} describes the testing capability of our proposed models using future long-baseline experiments such as DUNE, P2SO, T2HK, and T2HKK;
Section \ref{sec:g-2} presents an elaborative explanation on electron and muon $(g-2)$; finally, we conclude our results in Section \ref{sec:con}.
\section{$U(1)^{}_{L_e - L_{\alpha}}$ Model with Hybrid Seesaw Scenario} 
\label{sec:framework} 
We opt for a hybrid seesaw scenario in which neutrinos acquire their masses from a combination of the type-I and type-II frameworks. Here, type-I is realized with three right-handed neutrinos and the type-II contribution is attained with a scalar triplet of hypercharge +1. We discuss these in an extended SM scenario with
 additional $U(1)_{L_e - L_{\alpha}}$ gauge symmetry. An additional scalar singlet $S$ is introduced for the spontaneous breaking of new $U(1)$ symmetry. The particle contents of the two models are presented in Table. \ref{Table: model}.

\begin{table}[htpb]
\begin{center}
%================================================
\begin{tabular}{|c||c||c||c||c|}
	\hline
			Fields &  $SU(2)_L\times U(1)_Y$	& Model A: $U(1)_{L_{e}-L_{\tau}}$	& Model B: $U(1)_{L_{e}-L_{\mu}}$\\
	\hline
	\hline
	$\ell^{}_{e{\rm L}}, \ell^{}_{\mu{\rm L}}, \ell^{}_{\tau{\rm L}}$	& $(\textbf{2},  -1/2)$	&  $1,0,-1$	& $1,-1,0$\\ \hline
			 $e^{}_{\rm R}, \mu^{}_{\rm R}, \tau^{}_{\rm R}$							& $(\textbf{1},  -1)$	&  $1,0,-1$	& $1,-1,0$\\ \hline
			 $N_{e{\rm R}}, N_{\mu{\rm R}}, N_{\tau{\rm R}}$						& $(\textbf{1}, 0)$	&  $1,0,-1$	& $1,-1,0$\\
	\hline \hline
	  $H$							& $(\textbf{2},~ 1/2)$	&   $0$	& $0$\\ \hline  
		 $\Delta$							& $(\textbf{3},1)$	&   $1$	& $1$\\ \hline
			 $S$						& $(\textbf{1},~   0)$	&  $1$	& $1$\\  
	\hline
	\hline
\end{tabular}
%================================================
\caption{Fields and their charges in the chosen $SU(2)_L \times U(1)_Y \times U(1)_{L_{e}-L_{\alpha}}$ models.}
\label{Table: model}
\end{center}
\end{table}
 The  Lagrangian for the scalar sector takes the form 
 \begin{align}
\mathcal{L}_{\rm scalar} &= (D_\mu H)^\dagger (D^\mu H) + (D_\mu \Delta)^\dagger (D^\mu \Delta) +(D_\mu S)^\dagger (D^\mu S) - V,
\end{align}
where the scalar potential $V$ is provided as follows: 
\begin{align}
V &=  \mu^2_H  H^\dagger H + \lambda_H (H^\dagger H)^2 +  \mu^2_\Delta {\rm Tr} (\Delta^\dagger \Delta) +\lambda_\Delta {\rm Tr}(\Delta^\dagger \Delta\Delta^\dagger \Delta)  + \lambda^\prime_\Delta {\rm Tr}(\Delta^\dagger \Delta)^2 \nn\\   
      & + \mu^2_S  S^* S + \lambda_S (S^* S)^2 + \lambda_{HS} (H^\dagger H)(S^* S)+ \lambda_{H\Delta} (H^\dagger \Delta \Delta^\dagger H) + \lambda^\prime_{H\Delta} (H^\dagger H) {\rm Tr}(\Delta^\dagger \Delta)   
       \nonumber\\ & + \lambda_{\Delta S} {\rm Tr}(\Delta^\dagger \Delta) (S^* S)+ \frac{\lambda_{H\Delta S}}{2} \left[(H^T i\sigma_2\Delta^\dagger H)S + {\rm h.c}\right].   
\label{scalarpot}
\end{align} 
After spontaneous symmetry breaking, the scalar multiplets can be written as \\$S = \left(v_S + h_S + i A_S\right)/\sqrt{2}$,~~  $\Delta = \begin{pmatrix}
		 \Delta^+/\sqrt{2} & \Delta^{++}		\\
		 \Delta^0	& -\Delta^+/\sqrt{2}\\
	\end{pmatrix}$ and $H = \begin{pmatrix}
		 H^+ 		\\
		 H^0\\
	\end{pmatrix}$, with
\begin{eqnarray}
&& H^0 = \frac{v + h + iA_H}{\sqrt{2}},\nn\\
&& \Delta^0 = \frac{v_\Delta + h_\Delta + i A_\Delta}{\sqrt{2}},
\end{eqnarray}
where $v, v_{\Delta}$ and $v_S$ are the vacuum expectation values (VEVs) of the SM Higgs $H$ and the new scalars $\Delta$ and $S$, respectively.
 The covariant derivatives in the kinetic terms are expressed as 
\begin{eqnarray}
&&D_\mu H= \d_\mu H + ig \left(\frac{\sigma^a}{2}\cdot W^a_\mu\right) H + i \frac{g^\prime}{2} B_\mu H\;,\nn\\
&&D_\mu \Delta= \d_\mu \Delta + ig \left[\sum_{a=1}^3 \frac{\sigma^a}{2}W^a_\mu, \Delta\right] + i g^\prime B_\mu  \Delta + ig_{e \alpha} (Z_{e\alpha})_\mu \Delta\;,\nn \\
&&D_\mu S= \d_\mu S + ig_{e\alpha} (Z_{e\alpha})_\mu S. 
\end{eqnarray}
In the above, $\sigma_a$ with $a=1,2,3$ stands for the Pauli matrices. The new gauge boson associated with $U(1)$ symmetry   attains a mass $M_{Z_{e\alpha}} = g_{e\alpha} v_S$, with $g_{e\alpha}$ being the new gauge coupling strength. 

\subsection{Vacuum stability and Unitarity criteria}
The vacuum stability criteria \cite{Kannike:2012pe,Modak:2013jya} for the potential provided in Equation \ref{scalarpot} are
\begin{eqnarray}
    &&\lambda_H \ge 0, \lambda_\Delta+\lambda^\prime_\Delta \ge 0, \lambda_S \ge 0, \nn\\
    &&\overline{\lambda_1} = (\lambda_{H\Delta} + \lambda^\prime_{H\Delta}) + 2\sqrt{\lambda_H (\lambda_\Delta+\lambda^\prime_\Delta)}\ge 0,\nn\\ 
    &&\overline{\lambda_2} = \lambda_{HS} + 2\sqrt{\lambda_H \lambda_S} \ge 0,\nn\\
    &&\overline{\lambda_3} = \lambda_{\Delta S} +2\sqrt{(\lambda_\Delta + \lambda^\prime_\Delta)\lambda_S} \ge 0, \nn\\
    && \sqrt{\lambda_H (\lambda_{\Delta} + \lambda^\prime_{\Delta}) \lambda_S} + (\lambda_{H\Delta} + \lambda^\prime_{H\Delta})\sqrt{\lambda_S} + \lambda_{HS}\sqrt{\lambda_\Delta + \lambda^\prime_\Delta} + \lambda_{\Delta S} \sqrt{\lambda_H} \nn\\ &&+ \sqrt{2 \overline{\lambda_1} ~\overline{\lambda_2}~ \overline{\lambda_3}}\ge 0.
\end{eqnarray}
Applying the tree-level perturbative unitarity constraints on the scattering processes in the scalar sector, the zeroth partial wave amplitude \cite{Lee:1977eg} takes the form
\begin{equation}
a_0=\frac{1}{32\pi} \sqrt{\frac{4 p^{\rm CM}_i p^{\rm  CM}_f}{s}}\int_{-1}^{+1}  T_{2 \to 2}~d (\cos\theta).
\end{equation}
Here $p_{f,(i)}^{\rm CM}$ is the the centre of mass momentum of the final (initial) state, $s$ is the center of mass energy, and  $T_{2 \to 2}$ stands for the total amplitude of each $2\to2$ scattering processes. At high energies, the partial wave amplitudes, i.e. the quartic couplings, are constrained by the requirement of perturbative unitarity $\left| {\rm Re} (a_0)\right| \le \frac{1}{2}$, provided by
\begin{eqnarray}
&& \lambda_H, (\lambda_\Delta + \lambda^\prime_\Delta), \lambda_S \le \frac{4\pi}{3}, \nn\\
&& (\lambda_{H \Delta} + \lambda^\prime_{H\Delta}), \lambda_{HS}, \lambda_{\Delta S} \le 4\pi. 
\end{eqnarray}

\subsection{Scalar mass spectrum}
The multiple scalars of the model mix with each other based on their property. The minimisation conditions take the form \cite{Das:2016bir}
\begin{eqnarray}
    && \mu_H^2 = -\lambda_H v^2 - \lambda_{HS}\frac{v_S^2}{2} - \lambda_{H\Delta}\frac{v_\Delta^2}{2} - \lambda^\prime_{H\Delta}\frac{v_\Delta^2}{2} + \lambda_{H\Delta S}\frac{v_S v_\Delta}{2}, \nn\\
&& \mu_\Delta^2 = -\lambda_\Delta v_\Delta^2 - \lambda^\prime_\Delta v_\Delta^2 - \lambda_{H\Delta}\frac{v^2}{2} - \lambda^\prime_{H\Delta}\frac{v^2}{2} - \lambda_{\Delta S}\frac{v_S^2}{2} + \lambda_{H\Delta S}\frac{v_S v^2}{4 v_\Delta}, \nn\\  
&& \mu_S^2 = -\lambda_S v_S^2 - \lambda_{HS}\frac{v_S^2}{2}  - \lambda_{\Delta S}\frac{v_\Delta^2}{2} + \lambda_{H\Delta S}\frac{v_\Delta v^2}{4 v_S}.
\end{eqnarray}
The mass of the doubly charged scalar is given by
\begin{equation}
    M^2_{CC} = -\lambda_{H\Delta} \frac{v^2}{2} - \lambda_{\Delta}v^2_\Delta + \lambda_{H\Delta S} \frac{v_S v^2}{4v_\Delta}.
    \label{eqn_mcc}
\end{equation}
The mass matrix of singly charged scalars in the basis $(H^+, ~\Delta^+)^T$ takes the form
\begin{equation}
M^2_C = \frac{1}{2}(-\lambda_{H\Delta}v_\Delta + \lambda_{H\Delta S}v_S)\left( \begin{matrix}  v_\Delta & -\frac{v}{\sqrt{2}}~ \cr -\frac{v}{\sqrt{2}}  & ~\frac{v^2}{2v_\Delta}~ \end{matrix}\right).
\end{equation}
Upon diagonalisation, we have one mass eigenstate with mass zero (absorbed as Goldstone mode for $W^+$) and a massive charged scalar ($C_1$) with mass 
\begin{equation}
M^2_{C_1} = \frac{1}{4}\left(\lambda_{H\Delta S}\frac{v_S}{v_\Delta} - \lambda_{H\Delta }\right)(v^2 + 2v^2_\Delta).
\label{eqn_mc1}
\end{equation}
Moving on to the CP-odd sector, the mixing matrix in the basis $(A_H, ~A_\Delta ,~A_S)$ is
\begin{equation}
M^2_{O} =  \lambda_{H\Delta S}\left( \begin{matrix}  ~v_\Delta v_S & -\frac{v_S v}{2}~ & ~\frac{vv_\Delta}{2}~ \cr -\frac{v_S v}{2}~  & ~\frac{v_S v^2}{4v_\Delta}~ & ~-\frac{v^2}{4}~ \cr ~\frac{vv_\Delta}{2}~ & -\frac{v^2}{4}& ~\frac{v_\Delta v^2}{4v_S}~ \end{matrix}\right).
\end{equation}
After diagonalization, we obtain two massless eigen states (one Goldstone mode absorbed by the SM $Z$ boson and the other by the $Z_{e\alpha}$ gauge boson), and one massive CP-odd scalar ($O_1$) with mass
\begin{equation}
    M^2_{O_1} = \lambda_{H\Delta S} \left(\frac{v^2 v^2_\Delta + v^2 v^2_S + 4 v_\Delta^2 v_S}{4 v_\Delta v^2_S}\right).
    \label{eqn_mo1}
\end{equation}
Now, the mass matrix of CP-even scalars in the basis $(h ,~h_\Delta ,~h_S)$ takes the form 
\begin{equation}
M^2_E = \left( \begin{matrix} 2\lambda_H v^2 & (\lambda_{H\Delta}+\lambda^\prime_{H\Delta})vv_\Delta -\lambda_{H\Delta S} \frac{v_Sv}{2} & ~\lambda_{HS} vv_S -\lambda_{H\Delta S} \frac{vv_\Delta}{2}~ \cr (\lambda_{H\Delta}+\lambda^\prime_{H\Delta})vv_\Delta -\lambda_{H\Delta S} \frac{v_Sv}{2}  & ~2(\lambda_{\Delta}+\lambda^\prime_{\Delta})v^2_\Delta + \lambda_{H\Delta S} \frac{v_Sv^2}{4 v_\Delta} & ~\lambda_{\Delta S}v_Sv_\Delta - \lambda_{H\Delta S} \frac{v^2}{4}~ \cr ~\lambda_{HS} vv_S -\lambda_{H\Delta S} \frac{vv_\Delta}{2} & ~\lambda_{\Delta S}v_Sv_\Delta - \lambda_{H\Delta S} \frac{v^2}{4}  & ~2\lambda_S v_S^2 + \lambda_{H\Delta S} \frac{v^2v_\Delta}{4v_S} \end{matrix}\right).
\label{mat_even}
\end{equation}
%
%\subsubsection*{Physical scalar masses}
Assuming  the hierarchical nature of the VEVs, i.e.,  $v_\Delta \ll v < v_S$, the mass matrix of Equation \ref{mat_even} can be written in a simplified form as
\begin{equation}
M^2_E = \left( \begin{matrix} 2\lambda_H v^2 &  -\lambda_{H\Delta S} \frac{v_Sv}{2} & ~\lambda_{HS} vv_S~ \cr  -\lambda_{H\Delta S} \frac{v_Sv}{2}  & ~ \lambda_{H\Delta S} \frac{v_Sv^2}{4 v_\Delta} & ~\lambda_{\Delta S}v_Sv_\Delta - \lambda_{H\Delta S} \frac{v^2}{4}~ \cr ~\lambda_{HS} vv_S & ~\lambda_{\Delta S}v_Sv_\Delta - \lambda_{H\Delta S} \frac{v^2}{4}  & ~2\lambda_S v_S^2  \end{matrix}\right).
\end{equation}
Taking $\lambda_{HS} = 0$ and $\lambda_{\Delta S} = \lambda_{H\Delta S} ({v^2}/{4 v_\Delta v_S})$, the above matrix further simplifies to
\begin{equation}
M^2_E = \left( \begin{matrix} 2\lambda_H v^2 &  -\lambda_{H\Delta S} \frac{v_Sv}{2} & ~0~ \cr  -\lambda_{H\Delta S} \frac{v_Sv}{2}  & ~ \lambda_{H\Delta S} \frac{v_Sv^2}{4 v_\Delta} & ~0~ \cr ~0 & ~0  & ~2\lambda_S v_S^2  \end{matrix}\right).
\end{equation}
In our analysis, we consider the VEVs as $v = 246$ GeV, $v_\Delta \sim 0.1$ GeV and $v_S \sim 7$ TeV with the scalar couplings $\lambda_S = 0.1$, $\lambda_{H\Delta S} = 0.001$.
As the VEV
of S is large, the corresponding CP-even scalar is taken to be at high scale and is decoupled from the other scalar spectrum of the model. The above assumption is taken to simplify the expressions and allow for a simple transparent phenomenological study. Thus, we end up with three physical CP-even scalars; one of them is the observed Higgs at LHC, i.e., $H_1$ with mass $M_{H_1} = 125$ GeV, and the other two are the heavy physical scalars $H_2, H_3$ with masses $M_{H_2} \sim 1$ TeV and $M_{H_3} \sim 3$ TeV, respectively. Furthermore, using Equations \ref{eqn_mcc}, \ref{eqn_mc1}, and \ref{eqn_mo1}, we obtain $M_{CC} \simeq M_{C_1} \simeq M_{O_1} \sim 1$ TeV.

\section{Neutrino mass generation}

\subsection{Model-A}
\label{subsec:model A}
Here we consider the  $U(1)_{L_e-L_{\tau}}$ gauge
 extension of the SM to investigate the neutrino phenomenology. 
The particle content and respective charges for each model are provided in  Table. \ref{Table: model}. The Lagrangian for the leptonic sector is provided below,
\begin{eqnarray}
{\cal L}_{\rm Lepton} & 
 \supset
 &  -  \left( y^e_l \overline{\ell^{}_{e {\rm L}}} H e^{}_{\rm R} + y^\mu _l \overline{\ell^{}_{\mu _{\rm L}}} H \mu^{}_{\rm R} 
 + y^\tau _l \overline{\ell^{}_{\tau_{\rm L}}} H \tau^{}_{\rm R} \right)  -  \frac{1}{2} y^{}_\Delta \left( \overline{\ell_{\tau {\rm L}}^C} {\rm i}\sigma^{}_2  \Delta  \ell_{\mu{\rm L}}      
 + 
 \overline{\ell_{\mu {\rm L}}^C} {\rm i}\sigma^{}_2  \Delta  \ell_{\tau{\rm L}} \right) \nonumber \\ &&
   - \left( y^e_{\nu} \overline{\ell^{}_{e {\rm L}}} \widetilde{H} N^{}_{e {\rm R}} + y^\mu _{\nu} \overline{\ell^{}_{ \mu {\rm L}}} \widetilde{H} N^{}_{\mu {\rm R}} + y^\tau _{\nu} \overline{\ell^{}_{ \tau {\rm L }}} \widetilde{H} N^{}_{\tau {\rm R}} \right) - \frac{1}{2} y^{\mu \tau}_{S} \left( \overline{N^{\rm C}_{\mu{\rm R}}} N^{}_{\tau {\rm R}} + \overline{N^{\rm C}_{\tau{\rm R}}} N^{}_{\mu {\rm R}} \right) S    \nonumber \\ && -  \frac{1}{2} y^{e\mu}_{S} \left( \overline{N^{\rm C}_{e{\rm R}}} N^{}_{\mu {\rm R}} + \overline{N^{\rm C}_{\mu{\rm R}}} N^{}_{e {\rm R}} \right) S^*   -  \frac{1}{2}  m^{\mu\mu}_{\rm R} \overline{N^{\rm C}_{\mu {\rm R}}} N^{}_{\mu {\rm R}}   \nonumber \\ && -~ m^{e\tau}_{\rm R} \left(\overline{N^{\rm C}_{e {\rm R}}} N^{}_{\tau {\rm R}}  + \overline{N^{\rm C}_{\tau {\rm R}}} N^{}_{e {\rm R}}  \right) + {\rm h.c.}
    \label{eq:Yukawalep_le-ltau_A}
\end{eqnarray}
The Dirac and Majorana mass matrices for the left and right-handed neutrinos take the form  
  \begin{equation}
  \quad m^{}_{\rm D} =  \frac{v^{}}{\sqrt{2}} ~{\rm diag} \left (y^e_\nu ,~ y^\mu_\nu, ~y^\tau_\nu \right )  \; ,    
   \quad m^{}_{\rm R} = \left( \begin{matrix} 0 & y^{e\mu}_s {v^{}_S\over \sqrt{2}}  & ~m_R^{e\tau} \cr y^{e\mu}_s { v^{}_S\over \sqrt{2}}  & m^{\mu \mu}_R & y_s^{\mu\tau}\frac{v_s}{\sqrt2} \cr m_R^{e\tau} & y_s^{\mu\tau}\frac{v_s}{\sqrt2}  & 0 \end{matrix}\right) \;.      \label{eq:MD MR_le-ltau}
\end{equation}
For simplicity, we consider the couplings involved in the RH neutrino mass matrix to be of similar order, i.e., $ y^{ e \mu}_s \approx y^{\mu \tau}_s $, which is a justifiable assumption for the $U(1)_{L_e-L_\tau}$ model as the muon-type RH neutrino can couple to first- and third-generation neutrinos with similar strength.
Thus, the Majorana mass matrix takes the form,\begin{eqnarray}
     \quad m^{}_{\rm R} = \left( \begin{matrix} 0 & |y^{e\mu}_s| {v^{}_S\over \sqrt{2}} e^{i\phi_A} & ~m_R^{e\tau} \cr |y^{e\mu}_s |{ v^{}_S\over \sqrt{2}} e^{i\phi_A} & m^{\mu \mu}_R & |y^{e \mu}_s| {v^{}_S\over \sqrt{2}} e^{i \phi_A}  \cr m_R^{e\tau} & |y^{e \mu}_s| {v^{}_S\over \sqrt{2}} e^{i \phi_A}  & 0 \end{matrix}\right) \; ,
    \label{eq:MR_simpki}
    \end{eqnarray}
    where we have considered the coupling $y^{ e \mu}_s $ as complex with phase $\phi_A$.
  
  The  light neutrino mass matrix in this framework  receives contributions  from both type-I and type-II seesaw mechanisms. Type-I seesaw can provide the active neutrino mass as follows \cite{Gell-Mann:1979vob}:
\begin{equation}
  m_\nu =- m_D m_R^{-1} m^T_D\;,  
\end{equation}
  while the type-II contribution comes from a scalar triplet which takes the form 
\begin{eqnarray}
\quad      m^{}_{\rm L} = \frac{y^{}_\Delta v^{}_\Delta}{\sqrt{2}}  \left( \begin{matrix} 0 & ~~0~~ & 0 \cr 0 & 0 &  1 \cr  0 & 1 & 0  \end{matrix} \right).
     \label{eq:ML_le-ltau}
    \end{eqnarray}
Adding both these contributions, the expression for the light neutrino mass matrix becomes
\begin{eqnarray}
     \quad      m^{}_{\nu} =\left( \begin{matrix} a ~~& b&~~ c \cr b ~~& d &~~ e \cr c ~~& e &~~ f \end{matrix} \right) ,
     %     (7)
     \label{mnu-exp}
    \end{eqnarray}
    where,
\begin{equation*}
    a = \frac{Q^2 e^{2 i \phi_A}}{4P},
    ~b= \frac{-e^{i \phi_A} Q R}{2 \sqrt{2} P},
    ~c= \frac{2N-Q^2Me^{2 i \phi_A}}{4P},
    ~d= \frac{R^2}{2 P},
    ~e=\frac{y_{\Delta} v_{\Delta}}{\sqrt{2}} - \frac{R Q M}{2 \sqrt{2} P},
    ~f=\frac{Q^2 M^2 e^{2 i \phi_A}}{4 P},
    \end{equation*}
    with,
\begin{eqnarray*}
&&P=  \left[ - (m_R^{e \tau })^2 m_R^{\mu \mu} + m_R^{e \tau} v_S^2 (y_S^{e \mu })^2 e^{2i\phi_A}\right],
~ ~~Q= v v_s y_{\nu}^e y_{s}^{e \mu},
\nonumber\\
&&R= m_R^{e \tau} v_H y_{\nu}^{\mu}\;,~~M=\frac{y_{\nu}^{\tau}}{y_{\nu}^e},
~ ~~N=y_{\nu}^{\tau} v_H^2 y_{\nu}^e m_R^{e \tau} m_R^{\mu \mu}.
\end{eqnarray*}
After diagonalizing the neutrino mass matrix (\ref{mnu-exp}), it is possible to obtain the values of
various oscillation parameters.

\subsection{Model B}
\label{sec:model B}
{Next, we  consider $U(1)_{L_e-L_\mu}$ gauge symmetry as an extension of the SM. The charges of the particles under various gauge groups are indicated in the fourth column of Table \ref{Table: model}. The Lagrangian for leptonic sector for this model reads as follows:
 \begin{eqnarray}
    {\cal L}_{Lepton} & \supset & -  \left( y^e_l \overline{\ell^{}_{e {\rm L}}}{H} e{_R} + y^\mu _l \overline{\ell^{}_{\mu _{\rm L}}} H \mu^{}_{\rm R} + y^\tau _l \overline{\ell^{}_{\tau_{\rm L}}} H \tau^{}_{\rm R}  \right) -  \left( \frac{1}{2} y^{}_\Delta \left( \overline{\ell_{\tau {\rm L}}^C} {\rm i}\sigma^{}_2  \Delta  \ell_{\mu{\rm L}} + \overline{\ell_{\mu {\rm L}}^C} {\rm i}\sigma^{}_2  \Delta  \ell_{\tau{\rm L}} \right)   \right) \nonumber \\ &&
    -\left( y^e_{\nu} \overline{\ell^{}_{e {\rm L}}} \widetilde{H} N^{}_{e {\rm R}} + y^\mu _{\nu} \overline{\ell^{}_{ \mu {\rm L}}} \widetilde{H} N^{}_{\mu {\rm R}} + y^\tau _{\nu} \overline{\ell^{}_{ \tau {\rm L }}} \widetilde{H} N^{}_{\tau {\rm R}} \right)  -  \frac{1}{2} y^{\mu \tau }_{S} \left( \overline{N^{\rm C}_{\mu{\rm R}}} N^{}_{\tau {\rm R}} + \overline{N^{\rm C}_{\tau{\rm R}}} N^{}_{\mu {\rm R}} \right) S \nonumber \\ && -  \frac{1}{2} y^{e \tau}_{S} \left( \overline{N^{\rm C}_{e{\rm R}}} N^{}_{\tau {\rm R}} + \overline{N^{\rm C}_{\tau{\rm R}}} N^{}_{e {\rm R}} \right) S^*   -  \frac{1}{2}  m^{\tau \tau}_{\rm R} \overline{N^{\rm C}_{\tau {\rm R}}} N^{}_{\tau {\rm R}}  \nonumber \\&& -~ m^{e\mu}_{\rm R} \left(\overline{N^{\rm C}_{\mu {\rm R}}} N^{}_{e {\rm R}}  + \overline{N^{\rm C}_{e {\rm R}}} N^{}_{\mu {\rm R}} \right)+ h.c. \; .
    %     (5)
    \label{eq:Yukawalep_le-ltau_B}
    \end{eqnarray}
    }
For the above Lagrangian, the Dirac and Majorana matrices $m_D$ and $m_L$ are same as in the previous model mentioned in Equations (\ref{eq:MD MR_le-ltau}) and (\ref{eq:ML_le-ltau}). Further, assuming $ y^{e\tau}_s \approx y^{\mu\tau}_s$ for numerical simplification, the  $m_R $ matrix is found to have the form
\begin{eqnarray}
     \quad m^{}_{\rm R} = \left( \begin{matrix} 0 & m_R^{e \mu} & |y^{e\tau}_s| {v^{}_S\over \sqrt{2}} e^{i\phi_B} \cr m_R^{e \mu}  & 0 &|y^{e \tau}_s|  { v^{}_S\over \sqrt{2}}e^{i \phi_B} \cr |y^{e\tau}_s| {v^{}_S\over \sqrt{2}} e^{i\phi_B} & |y^{e \tau}_s|  { v^{}_S\over \sqrt{2}}e^{i \phi_B} & m_R^{\tau \tau} \end{matrix}\right) \;,
\label{mr-simply}
\end{eqnarray}
where, $\phi_B$ represents the phase of $y_s^{e \tau}$. Now, analogous to Equation (\ref{mnu-exp}),  the expression of active neutrino mass matrix in type I+II for model B can be derived accordingly. 

Appropriate values of Yukawa couplings and VEVs of the scalar particles provide the required values  for neutrino oscillation parameters, which is discussed in the next section.

%%%%%%%%%%%%%%%%%%%%%%%%%%%%%%%%%%%%%%%%%%%%%%%%%%%%%%%%%%%%%%%%%%%
\section{Numerical Analysis}
\label{sec:numerical}
In this work, we perform numerical analysis by considering experimental data on various neutrino oscillation parameters at 3$\sigma$ interval of NuFit v5.2 \cite{Esteban:2020cvm}, as follows:
\begin{align}
& \Delta m^2_{\rm 31}=[2.427, 2.590]\times 10^{-3}\ {\rm eV}^2,\
\Delta m^2_{\rm 21}=[6.82, 8.03]\times 10^{-5}\ {\rm eV}^2,\nn\\
&\sin^2\theta_{13}=[0.02052, 0.02398],\ 
\sin^2\theta_{23}=[0.408, 0.603],\ 
\sin^2\theta_{12}=[0.270, 0.341]\;. \label{eq:mix}
%%% 
\end{align} 

\begin{table}[htb]
 \begin{center}
 \begin{tabular}{|c||c|}
 
 \hline
 \textbf{~~~~~~~~~~Model A} ~~~~~~~~~~~~~\hspace*{0.125 true cm} & ~~~~~~~~~~~~\textbf{Model B~~~~~~~~~~~~} \\
 \hline
 \end{tabular}
 \begin{tabular}{|c|c||c|c|}
 \hline
 Parameters  &  Range &  Parameters & Range \\
 \hline
 $y_{\nu}^{\beta}$   & $[1, 20] \times 10^{-7} $  &  $  y_{\nu}^{\beta} $    &  $[1,20] \times 10^{-7} $   \\ 
  $y_s^{e \mu}  $  &  $[0.01,0.1]  $    & $y_s^{e \tau}$ &   $[0.015,0.08] $   \\
  $y_{\Delta}$   &  $[20,45] \times 10^{-11} $   & $y_{\Delta}$   &  $[10, 30] \times 10^{-11} $  \\
  $m_R^{e \tau}$ (in TeV) &  $[0.5, 10 ] $  & $ m_R^{e \mu} $ (in TeV)&  $[0.5,10] $  \\
  $m_R^{\mu \mu} $ (in TeV)  &  $[1,10] $   & $m_R^{\tau \tau}$ (in TeV) &  $[1,10]$  \\
  $\phi_A$   &   $ [0, 2\pi] $   & $\phi_B$   &   $  [0, 2\pi]  $   \\
 \hline  
 \end{tabular}
 \caption{Ranges of parameters in model A (left column) and model B (right column) over which the scan was performed.}
 \label{table: best fit} 
 \end{center}
\end{table}  

Here, numerical diagonalization of the light neutrino mass matrix for both model A and model B is performed through $U_\nu^\dagger {M_\nu}U_\nu= {\rm diag}(m_1^2, m_2^2, m_3^2)$, where,  ${M_\nu}=m_\nu m_\nu^\dagger$ and $U_\nu$ is an unitary matrix. Thus, the neutrino mixing angles can be extracted using the standard relations \cite{Hochmuth:2007wq,Frampton:2004ud}
\begin{eqnarray}
\sin^2 \theta_{13}= |U_{13}|^2\;,~~~~\sin^2 \theta_{12}= \frac{|U_{12}|^2}{1-|U_{13}|^2}\;,~~~~~\sin^2 \theta_{23}= \frac{|U_{23}|^2}{1-|U_{13}|^2}\;.
\label{mixangles}
\end{eqnarray}

In order to demonstrate the current neutrino oscillation data,  we perform a random scan over the  ranges of the model parameter values provided in Table \ref{table: best fit}, where $\beta$ represents $e, \mu$ and $\tau$. We compute $\Delta m_{21}^2$, $\Delta m_{31}^2$ and $\delta_{CP}$ along with three mixing angles for all generated sets of model parameters. We utilize the $3\sigma$ bound on the neutrino mixing parameters \cite{Esteban:2020cvm} to obtain the consistent parameter space. Below, we illustrate our results with suitable figures.

\begin{figure}[htb!]
\begin{center}
\includegraphics[height=55mm,width=75mm]{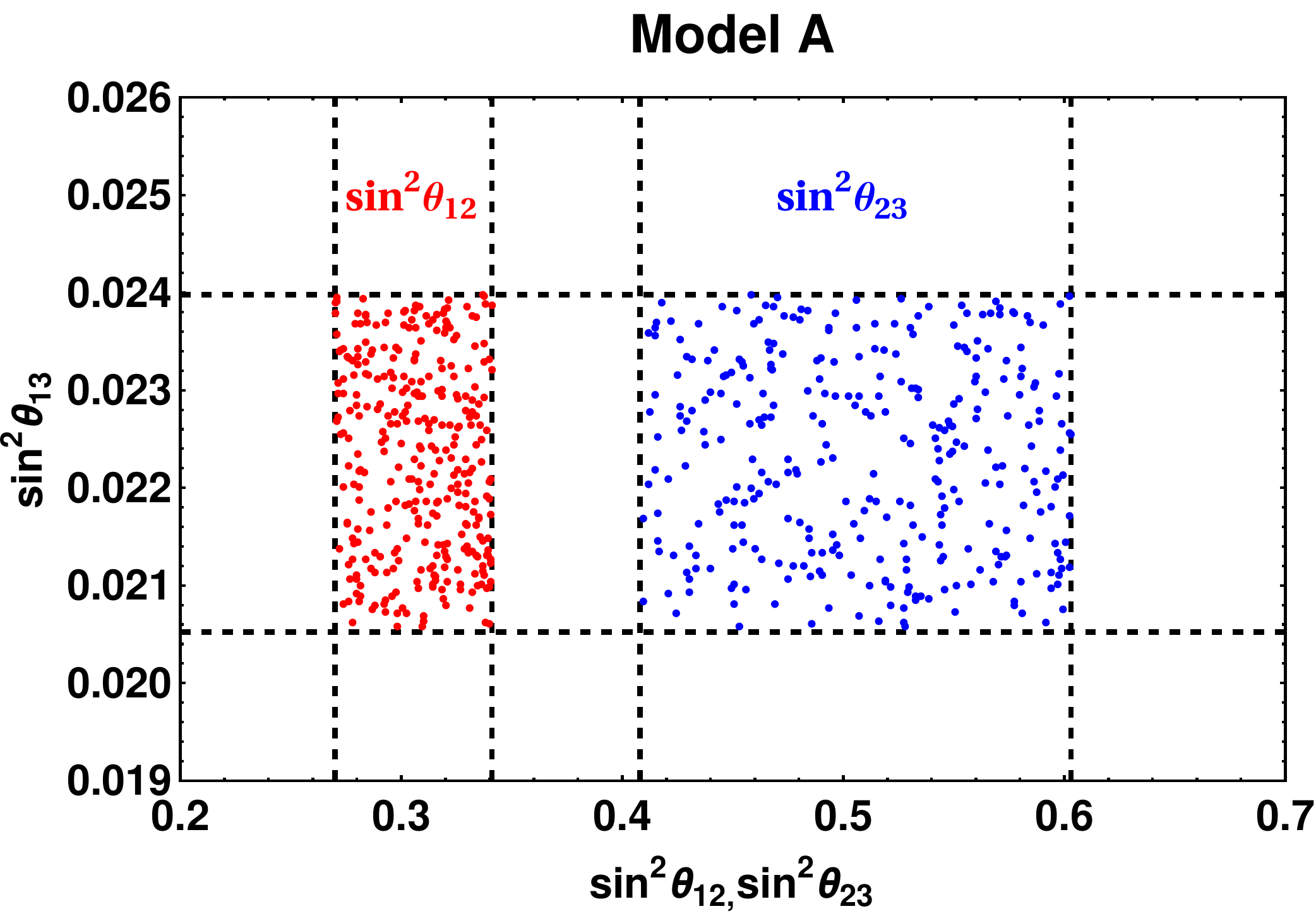}
\hspace*{0.2 true cm}
\includegraphics[height=55mm,width=75mm]{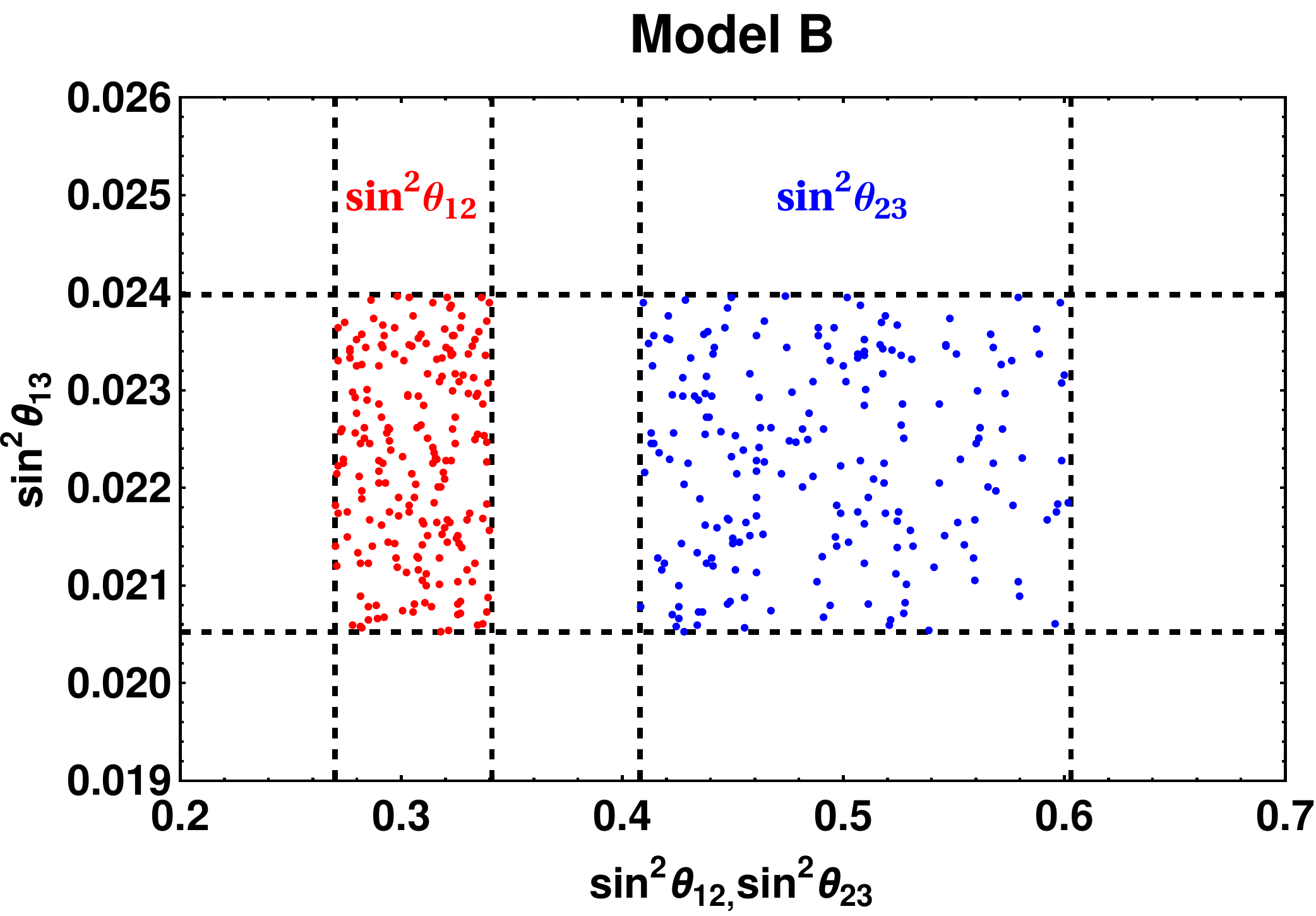}\\
\includegraphics[height=50mm,width=75mm]{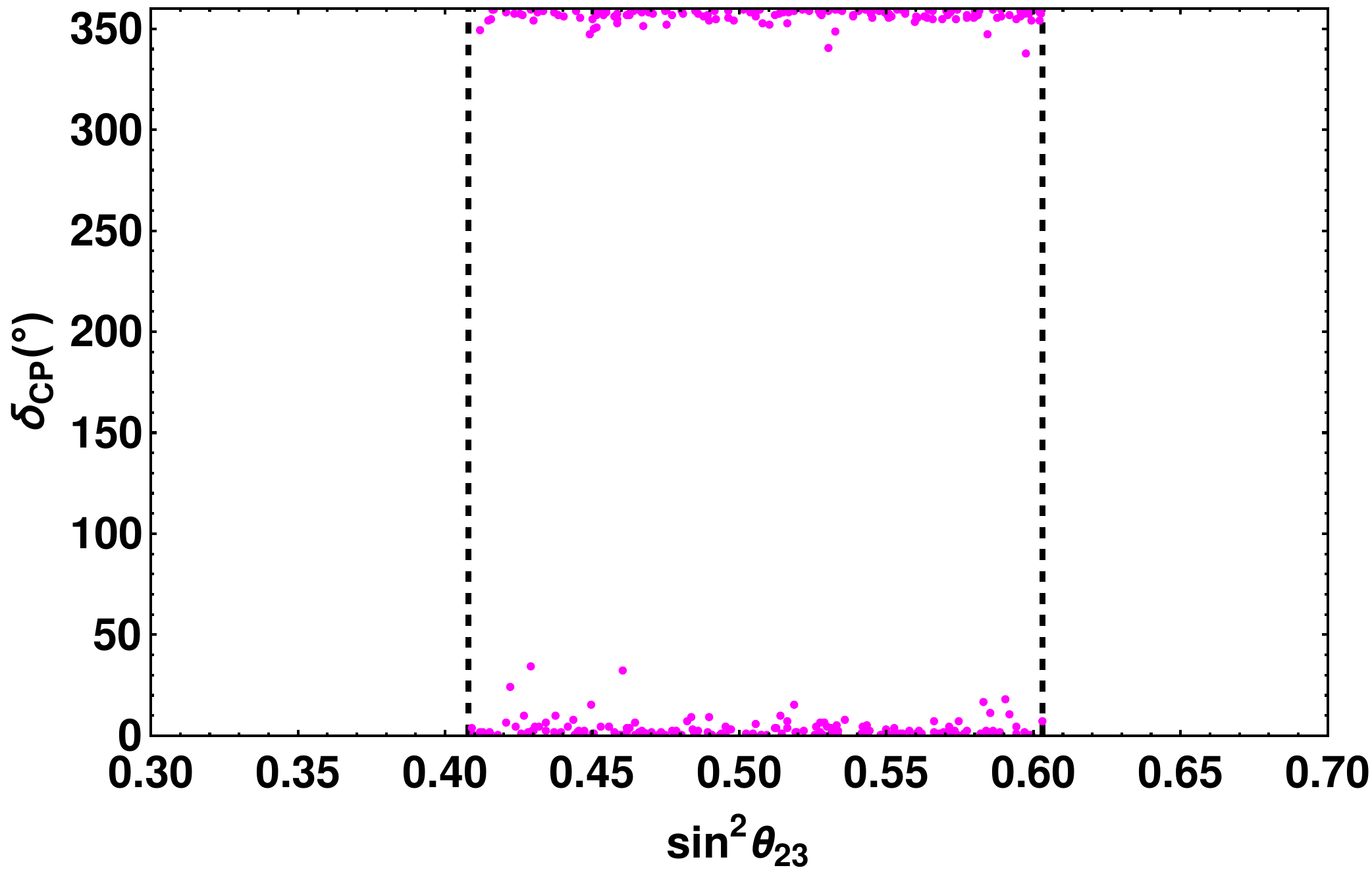}
\hspace*{0.2 true cm}
\includegraphics[height=50mm,width=75mm]{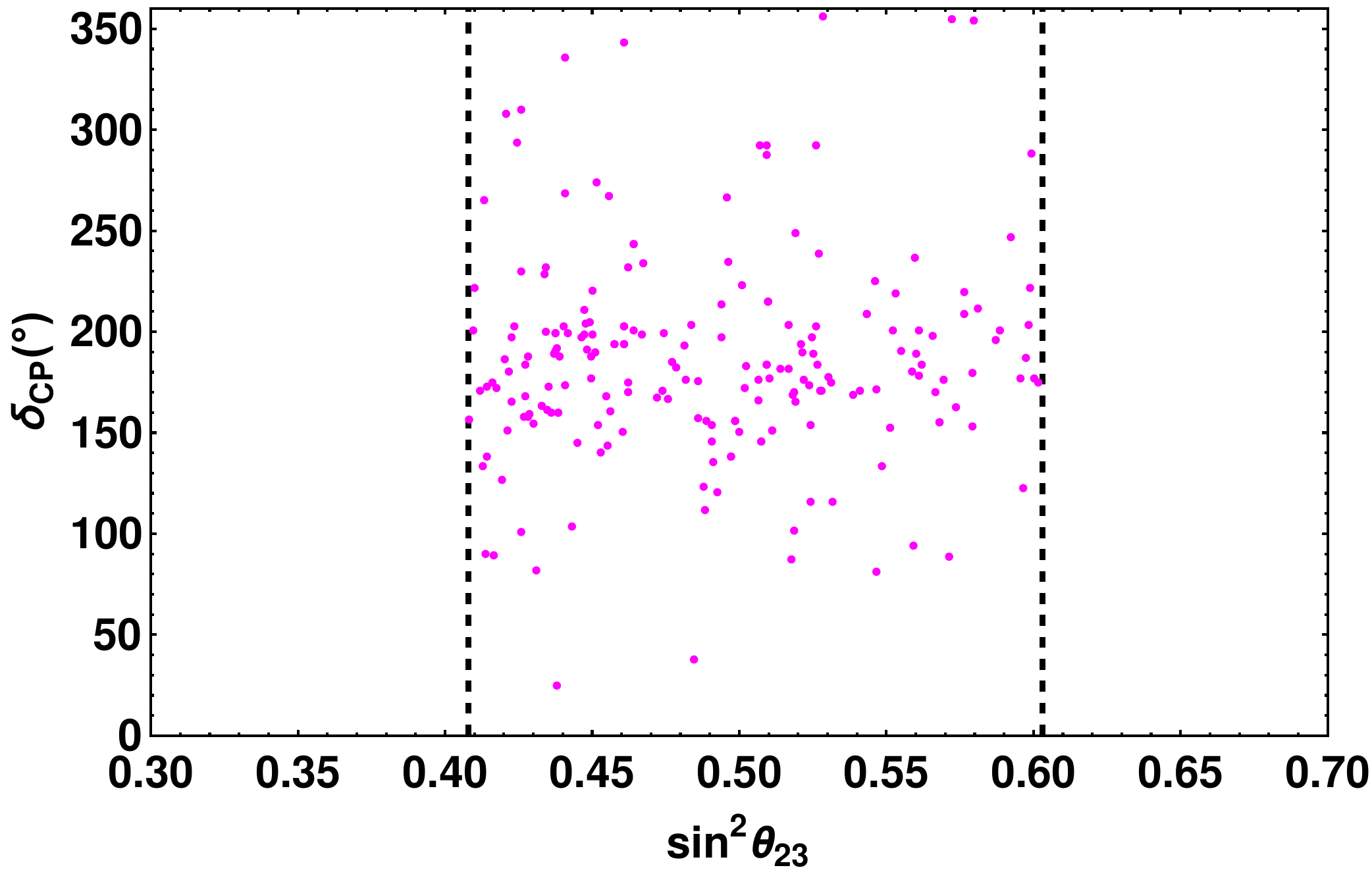}\\
\includegraphics[height=50mm,width=75mm]{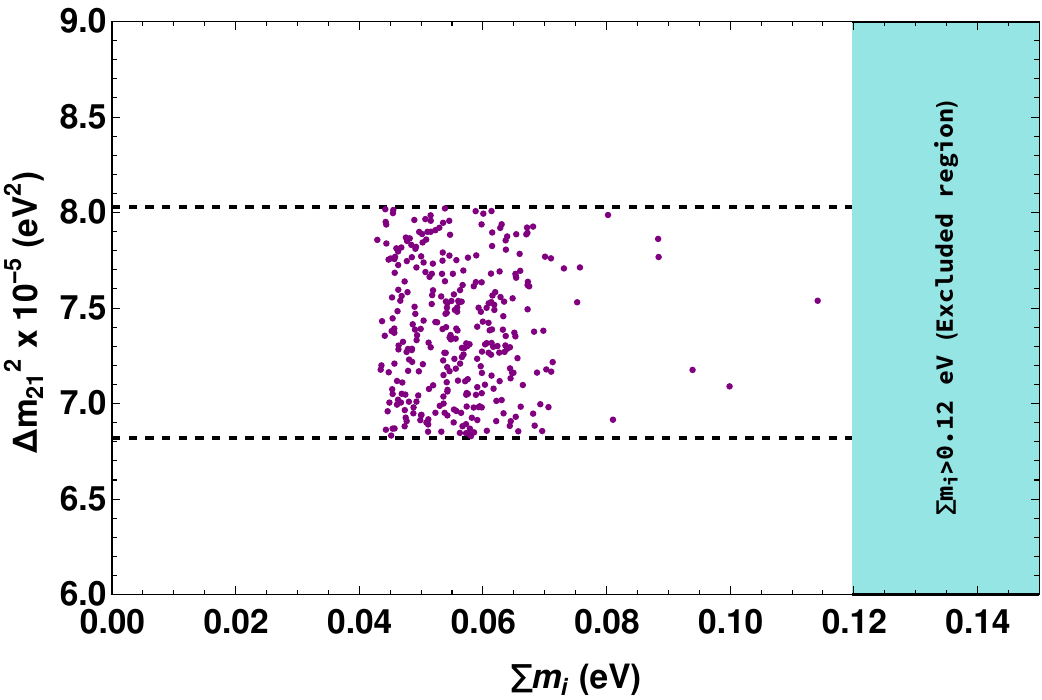}
\hspace*{0.2 true cm}
\includegraphics[height=50mm,width=75mm]{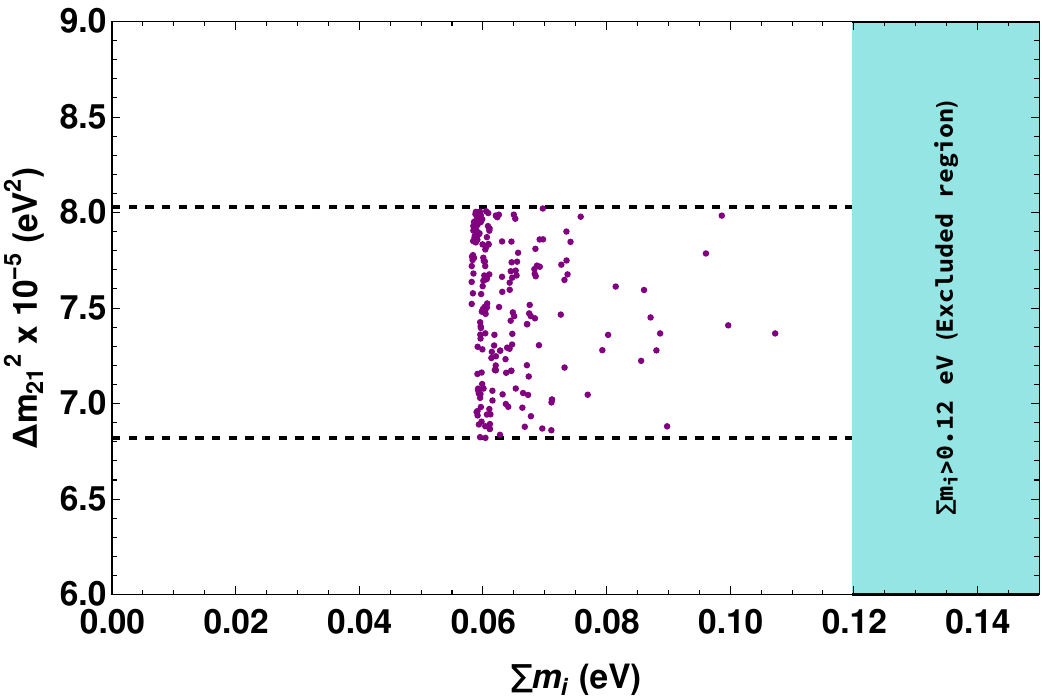}
\caption{Left (right) column signifies the obtained range of neutrino oscillation parameters  for model A (B). The top row represents the correlation plot between $\sin^2 \theta_{13}$ and $(\sin^2 \theta_{12},\sin^2 \theta_{23})$, whereas the plots in the middle row show the variation of $\delta_{\rm CP}$ with respect to $\sin^2 \theta_{23}$ and the bottom row depicts the variation in the sum of the active neutrino masses with $\Delta m_{21}^2$ for both models.}.
\label{all-param}
\end{center}
\end{figure}

Fig. \ref{all-param} shows the correlation plots for different neutrino oscillation parameters for the two models; the left column is for model A  and the right column is for model B. The gridlines in each plot represent the $3 \sigma$ allowed parameter space for the corresponding oscillation parameters. The top row depicts the allowed parameter space of $\sin^2 \theta_{13}$  and $(\sin^2 \theta_{12},\sin^2 \theta_{23})$ for both models. From these plots, we can conclude that both models
satisfy the full $3 \sigma$ region of NuFit v5.2 data for the plane  $\sin^2 \theta_{13}-\sin^2 \theta_{12} ~(\sin^2 \theta_{23})$; hence, they cannot provide any strong correlation for these oscillation parameters.
The middle row shows the variation of $\delta_{CP}$ with respect to $\sin^2 \theta_{23}$ for both the models. Unlike the
previous two cases, this row shows a strong constraint on the $\delta_{CP}$ value for model A. In
Model A, $\delta_{CP}$ is strongly biased towards CP-conserving values, specifically, $0^{\circ}$ and $360^{\circ}$; on
the other hand, in Model B $\delta_{CP}$ spans nearly the entire range from $20^{\circ}$ to $350^{\circ}$. Additionally,
the bottom row of Figure \ref{all-param} shows the variation in the sum of the active neutrino masses
($\sum_i  m_i$) with $\Delta m_{21}^2$; the left plot is for model A and the right plot is for model B. Model A
estimates the sum of active neutrino masses to be within the range of 0.043 eV to 0.114 eV,
while model B provides a range of 0.058 eV to 0.107 eV, keeping the $\Delta m_{21}^2$ value fully within
the $3 \sigma$ range for both models. Similarly, for $\Delta m_{31}^2$ we can check that the range is within the
full $3 \sigma$ region of NuFit v5.2 for both models.

From  the plots of Figure \ref{all-param}, we can conclude that among all six oscillation parameters,  model A only provides a strong bound on $\delta_{\rm CP}$. Thus, to obtain a complete picture of testability of the models in the upcoming long-baseline experiments, we   illustrate the variation of $\delta_{CP}^{\rm true}$ with $\theta_{23}^{\rm true}$, taking the NuFit v5.2 oscillation parameters as the true values in the following section.

%%%%%%%%%%%%%%%%%%%%
\section{Testing of the models in the upcoming long-baseline experiments}
\label{test-model}
One of the best ways to verify the predictions of proposed models such as ours is to probe them using long-baseline neutrino experiments. Several outstanding long-baseline
neutrino experiments are currently pending around the world, e.g., DUNE, P2SO, T2HK
and T2HKK etc. In this section, we show how our proposed models can be probed in these
future experiments. 

\subsection*{Experimental details}
DUNE (Deep Underground Neutrino Experiment)  is an upcoming long-baseline neutrino experiment with a baseline of  1300 km. The far detector of DUNE consists of four modules of LArTPC (Liquid Argon Time Projection Chamber) detectors, each with a volume of 10 kt.  The specifications regarding backgrounds, systematic errors, etc., are taken from \cite{DUNE:2021cuw}. In our calculation, we take five years of run-time in  neutrino mode and five years in anti-neutrino mode, with a POT (Proton on Target) of $1.1 \times 10^{21}$. In addition, we take a systematic error of $2 \% ~(5 \%)$ in the $\nu_{e} ~(\nu_{\mu})$ appearance (disappearance) signals.

Another highly promising long-baseline experiment is Protvino to Super ORCA (P2SO). It has a baseline of 2595 km from the neutrino source. The Super-ORCA detector   is a ten times more instrumented version of ORCA detector. For our simulation, we take three years of runtime in  neutrino mode and three years in anti-neutrino mode with a beam power of 450 KW, corresponding to $4 \times 10^{20}$ POT. For the P2SO configuration, we take a systematic error of $5 \% ~(5 \%)$ in the $\nu_{e} ~(\nu_{\mu})$ appearance (disappearance) signals and use the same configuration as used in  \cite{Singha:2022btw}.

The T2HK (Tokai to Hyper Kamiokande) experiment is an upcoming long-baseline neutrino experiment at
Japan. It  will have two large water Cherenkov (WC) far detectors. Each of the detectors has a fiducial volume of 187 kt. The distance from the neutrino source
(J-PARC) to Hyper-Kamiokande (HK) is around 295 km. We take the neutrino beam for T2HK at 1.3 MW with a POT of $2.7 \times 10^{22}$.
For T2HK, we use the configuration provided in  \cite{Hyper-Kamiokande:2016srs}. In our simulation, we take
the runtime of 10 years with an equal ratio (five years for neutrino mode and five years for
antineutrino mode). For T2HK configuration, we take a systematic error $4.71 \% ~(4.13 \%)$  in the $\nu_{e} ~(\nu_{\mu})$ appearance (disappearance) signals.

T2HKK (Tokai to Hyper Kamiokande and Korea) is the extended version of the T2HK experiment, in which one of the two water Cherenkov detectors from HK
will be shifted to Korea, where it will be located 1100 km away from the J-PARC source.
 For T2HKK, we
use an off-axis flux of 295 km $2.5^{\circ}$ and an off-axis flux of 1100 km $1.5^{\circ}$. Similar to
the T2HK configuration \cite{Hyper-Kamiokande:2018ofw}, for T2HKK, we take a systematic error $3.8 \% ~(3.8 \%)$  in the $\nu_{e} ~(\nu_{\mu})$ appearance (disappearance) signals. The runtime for this experiment is five years of neutrino mode and five years of anti-neutrino mode.
 \subsection*{Simulation details}
We use the General Long Baseline Experiment Simulator (GLoBES) \cite{Huber:2004ka, Huber:2007ji} package to simulate future experiments. For the sensitivity calculation, we use the Poisson log-likelihood formula:
 \begin{equation}
    \chi^2_{\rm stat} = 2 \sum_{i=1}^n  \left[ N_i^{ \rm test} - N_i^{\rm true} - N_i^{\rm true} ~\rm {ln} \left(\frac{N_i^{\rm test}}{N_i^{\rm true}} \right) \right] \;,
\end{equation}
where $N_{i}^{\rm true ~(\rm test)}$ is the total event rate from the true (test) spectrum, and `$i$' is the number of energy bins. For the purposes of our simulations, we take the values of the oscillation parameters from NuFit v5.2 (listed in Table \ref{NuFit}) \cite{Esteban:2020cvm} as the true values.
\begin{table}[]
    \centering
    \begin{tabular}{|c||c|}
    \hline
    \hline
       Oscillation parameters  & NuFit v5.2 \cite{Esteban:2020cvm} \\
       \hline
       \hline
        $\theta_{13}$ & $8.58^{\circ}$\\
        \hline
        \hline
        $\theta_{12}$ & $33.41^{\circ}$ \\
        \hline
        \hline
        $\theta_{23}$  & varied within $[37^{\circ}-53^{\circ}]$\\
        \hline
        \hline
        $\Delta m_{21}^2 \rm {~(eV^2)}$  &  $7.410 \times 10^{-5}$\\
        \hline
        \hline
        $\Delta m_{31}^2 \rm {~(eV^2)}$  & 
        $2.507 \times 10^{-3}$\\
        \hline
        \hline
        $\delta_{\rm {CP}}$  &  varied within [$0^{\circ}-360^{\circ}$]\\
        \hline
        \hline
     \end{tabular}
    \caption{Oscillation parameters with their NuFit v5.2 \cite{Esteban:2020cvm} best fit values and the required ranges used in Sec. \ref{test-model}}
    \label{NuFit}
\end{table}

\subsection*{Results}
\begin{figure}
    \centering
    \includegraphics[height=65mm, width=75mm]{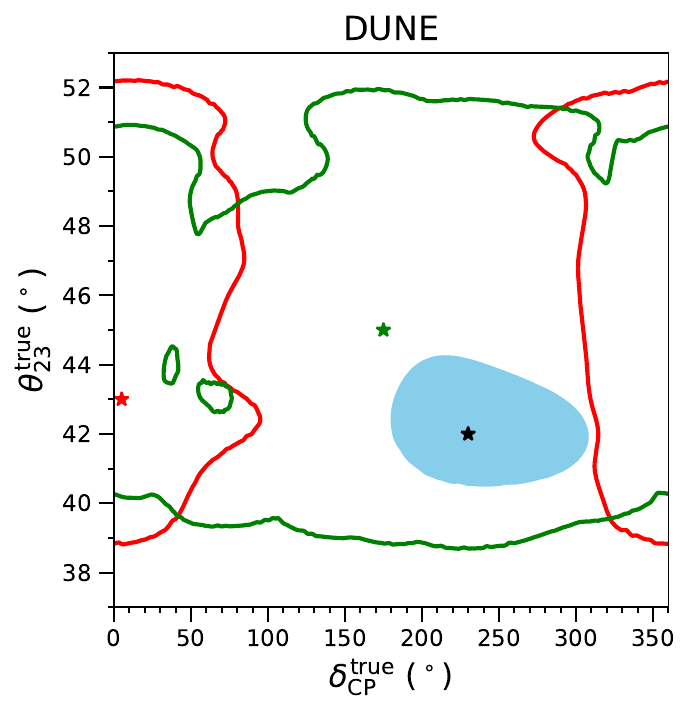}
    \includegraphics[height=65mm, width=75mm]{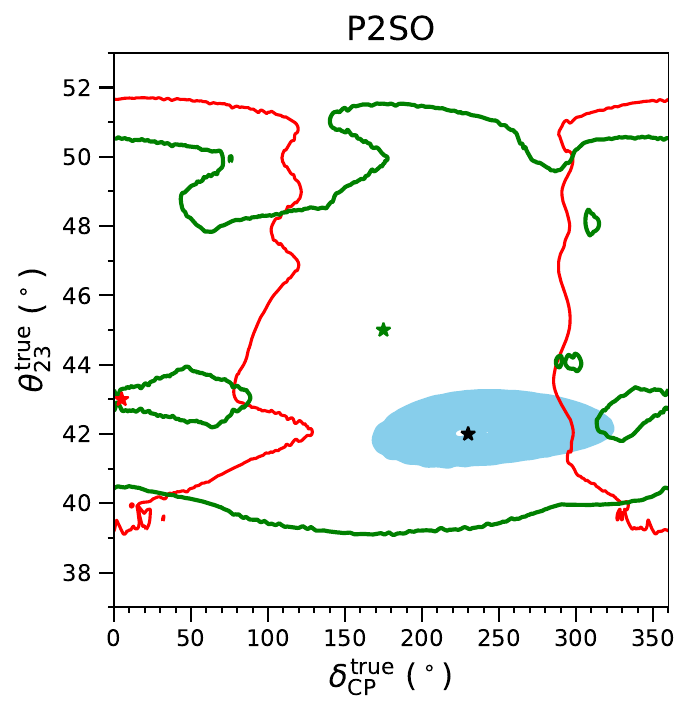}\\
    \includegraphics[height=65mm, width=75mm]{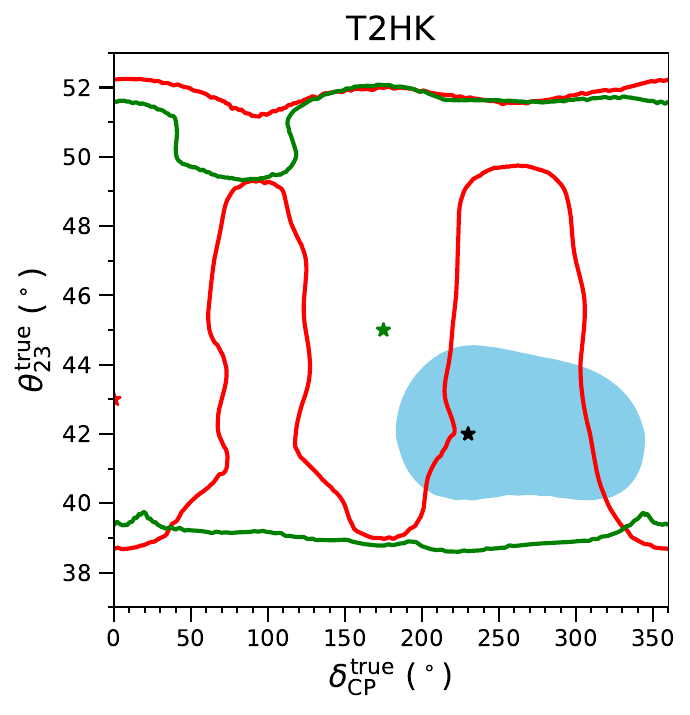}
    \includegraphics[height=65mm, width=75mm]{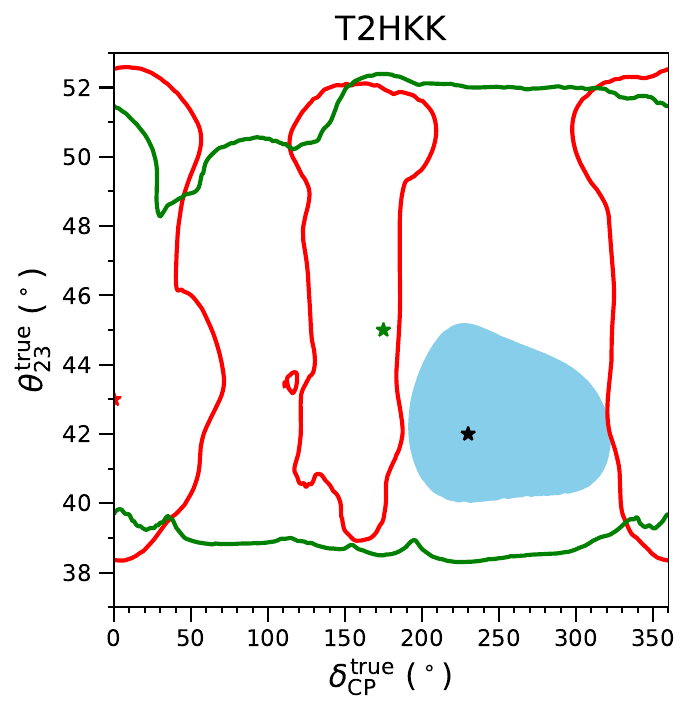}\\
    \includegraphics[height=40mm, width=70mm]{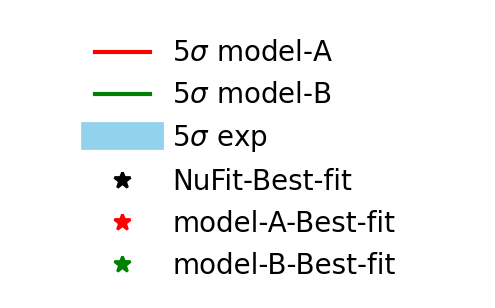}
    \caption{Sensitivities of DUNE, P2SO, T2HK and T2HKK in the $\theta_{23}^{\rm true}-\delta_{\rm CP}^{\rm true}$ plane used in probing the proposed models. The 
    left (right) plot in the upper panel shows the sensitivity of DUNE (P2SO) for both models. The
    left (right) plot in the lower panel shows the sensitivity of T2HK (T2HKK) for both  models.
The levels of each plot are mentioned in the legend below.}
    \label{test}
\end{figure}

In this subsection, we discuss the results of testing our proposed models in the future neutrino experiments DUNE, P2SO, T2HK, and T2HKK. Figure \ref{test} shows the parameter space of our proposed models with future experiments in the plane of $\theta_{23}^{\rm true}$ and $\delta_{\rm CP}^{\rm true}$. In each plot in Figure \ref{test}, the red and green curves represent the allowed parameter spaces of model A and model B, respectively, whereas the skyblue contours depict the parameter space of the different experiments with 
 the NuFit v5.2 best-fit values as the true values of
the oscillation parameters. All of the curves and contours are in $5 \sigma$ C.L.. The red, green, and black stars represent the best-fit values of models A, B, and NuFit v5.2, respectively. We take current NuFit v5.2 values as the true parameters for these plots to test our proposed proposed model's data. 

The left plot in the upper panel shows how the DUNE experiment can probe our models. The plot shows that the $5 \sigma$ allowed region of model A is incompatible with the $5 \sigma$ parameter space of DUNE. In contrast, the $5 \sigma$ allowed region of model B shows compatibility with the allowed parameter space of DUNE. Thus, DUNE can probe model B within its $5\sigma$ allowed region, assuming that the NuFit v5.2 oscillation parameters remain true.
However, when considering the  best-fit values,it can be seen that the values of both
models are outside the allowed $5 \sigma$ range of DUNE. Proceeding further to the right plot in the top panel, it can be observed that the $5 \sigma$ allowed region of both models A and B are compatible with the $5 \sigma$ parameter space of the P2SO experiment. Thus, our models can be probed by the P2SO experiment at $5 \sigma$ C.L. From the left plot of bottom panel, it can be seen that the $5 \sigma$ allowed ranges of models A and B are very much compatible with the $5 \sigma$ parameter space of T2HK, as some regions of the red, green, and skyblue curves intersect. From this plot, we can conclude that if the current NuFit v5.2 oscillation data remain true in future experiments, then T2HK will be able to probe our models at $5 \sigma$ C.L.

Finally, the right plot in the lower panel plots $\theta_{23}^{\rm true}$ with $\delta_{\rm CP}^{\rm true}$ for the T2HKK experiment. In this plot, it can be seen that the $5 \sigma$ allowed region of models A does not overlap with the $5 \sigma$ parameter space of T2HKK experiment. However, this is not the case for model B, where the allowed parameter space is compatible with T2HKK's $5\sigma$ C.L..  
Thus, we can conclude that if the NuFit v5.2 data remain as the true oscillation parameters, then the
T2HKK experiment will have the potential to probe Model B but not Model A.

%%%%%%%%%%%%%%%%%%%%%%%%%%%%%%%%
%%%%%%%%%%%%%%%%%
\section{ Electron and Muon $(g-2)$}
\label{sec:g-2}

\subsection{Electron $(g-2)$}
\label{sub:eg-2}

\begin{figure}
\begin{center}
\begin{tikzpicture}
\begin{feynman}
\vertex (a);
\vertex [left = of a] (b);
\vertex[below right = of a] (c);
\vertex[above right = of c] (d);
\vertex[right = of d]  (e);
\vertex[below = of c] (f);
%\vertex[above = of c] (g);

\diagram [line width=1.5pt]{
       (b) -- [black,fermion, edge label=$l^-$] (a) -- [black,fermion, edge label'=$l^-$]  (c);
       (c)   -- [black,fermion, edge label'=$l^-$] (d);
       (c) -- [red,boson,edge label=$\gamma$] (f);
     (a) -- [ violet,boson, half left, edge label=$Z_{e \alpha}$] (d);
     (d) --[black,fermion, edge label=$l^-$]  (e);
%     (e) -- [fermion]  (f);
};
\end{feynman}
\end{tikzpicture}
\hspace{0.9 true cm}
\begin{tikzpicture}
\begin{feynman}
\vertex (a);
\vertex [left = of a] (b);
\vertex[below right = of a] (c);
\vertex[above right = of c] (d);
\vertex[right = of d]  (e);
\vertex[below = of c] (f);

\diagram [line width=1.5pt]{
       (b) -- [black,fermion, edge label=$\mu^-$] (a) -- [black,fermion, edge label'=$\tau^+$]  (c);
       (c)   -- [black,fermion, edge label'=$\tau^+$] (d);
       (c) -- [red,boson,edge label=$\gamma$] (f);
     (a) -- [ violet,scalar, half left, edge label=$\Delta^{++}$] (d);
     (d) --[black,fermion, edge label=$\mu^-$]  (e);
};
\end{feynman}
\end{tikzpicture}
\caption{Possible Feynman diagrams for electron and muon $(g-2)$ calculation. For model A, only the left panel is applicable with $l^-=e^-$. For model B, both the left and right panels are present with $l^-= e^-,\mu^-$.}
\label{feyn}
\end{center}
\end{figure}

One of the open questions in particle physics is how to deal with electron and muon anomalous magnetic moments. Both of our models contribute to electron $(g-2)$  due to the fact that flavor symmetry has electron involvement i.e.,  $U(1)_{L_e-L_\tau}$ (model A) and $U(1)_{L_e-L_\mu}$ (model B). Further, the anomalous magnetic moment for electron is not accurately measured, i.e., its value is negative due to improved measurement of the fine-structure constant $\alpha_{\rm em}$ from  Cesium atoms \cite{Parker:2018vye} along with updated theoretical calculations \cite{Aoyama:2017uqe}, while    from the recent measurement of $\alpha_{\rm em}$ with Rubidium atoms \cite{Morel:2020dww} its value is found to be positive. The Cesium atom measurement of electron $(g-2)$ is \cite{Parker:2018vye}
\begin{equation}
\Delta a_e = (-8.7 \pm 3.6 ) \times 10^{-13}\;,
\end{equation}
whereas the Rubidium atom measurement provides the value of electron anomalous magnetic moment as \cite{Morel:2020dww} 
\begin{equation}
\Delta a _e = (4.8 \pm 3.0) \times 10^{-13}. 
\end{equation}

Numerous approaches have been attempted for explaining electron $(g-2) $ \cite{Giudice:2012ms,NA64:2021xzo}. Our obtained result for the anomalous magnetic moment is within the experimental range provided by Rubidium atoms.
 Below, we illustrate the discussion of electron $(g-2)$  using models A and B.

In model A, the contribution for $\Delta a_e$ comes from the gauge boson $Z_{e \tau}$ only, as there is no interaction between the electron and scalar triplet $\Delta $. Left panel of Figure \ref{feyn}  shows the possible Feynman diagram for model A with $l^-=e^-$. The interaction Lagrangian in this model contributing to the electron $(g-2)$ is provided as follows \cite{Moore:1984eg}:
\begin{equation}
\mathcal{L}= g_{e \tau} \bar{e} \gamma^\mu e Z_{e \tau} + h.c.\;, 
\end{equation}
where $g_{e \tau}$ and $Z_{e \tau}$ are the gauge coupling and gauge boson for $U(1)_{L_e-L_\tau}$ symmetry respectively. Thus, we can obtain the expression for $a^{Z_{e \tau}} _e $  as \cite{Moore:1984eg}
\begin{equation}
a^{Z_{e \tau}}_e = \frac{ g^2_{e \tau}}{4 \pi^2 } \int_0^1 \frac{x^2 (1-x)}{x^2 + \frac{m^2_{Z_{e \tau}}}{m^2_e }(1-x)}dx,
\end{equation}
where $x$ is the Feynman parameter. After simplification $(m_{Z_{e \tau}} \gg m_e)$, we obtain the relation
\begin{equation}
a^{Z_{e \tau}} _e=  g^2_{e \tau} \frac{m^2_e }{12 \pi ^2 m^2_{Z_{e \tau}} }\;.
\end{equation}

\begin{figure}
\centering
\includegraphics[height=55mm,width=75mm]{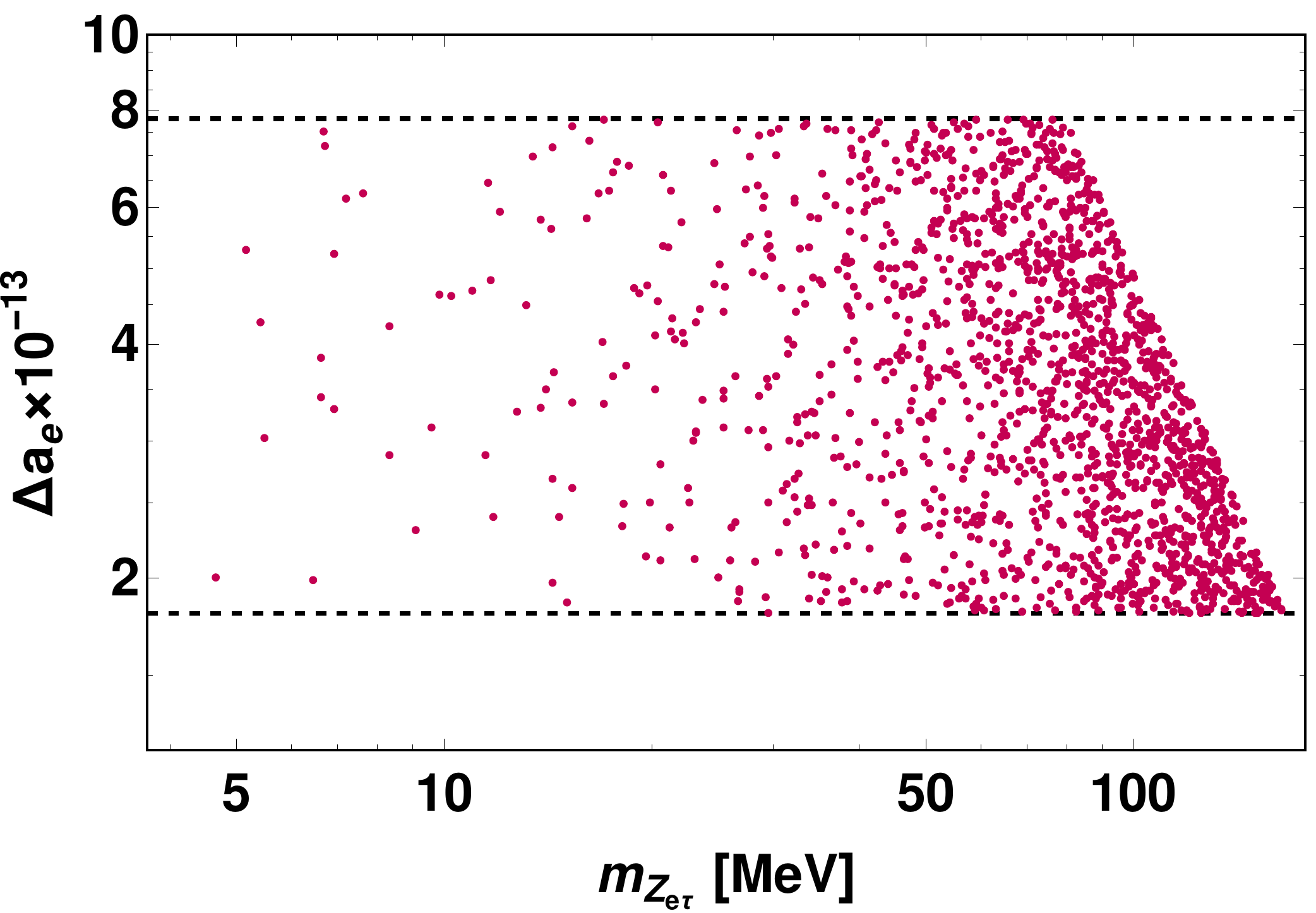}
\caption{Variation of electron anomalous magnetic moment with respect to the new gauge boson mass $m_{Z_{e \tau}}$ for model A.}
\label{dae-mz}
\end{figure}

Figure \ref{dae-mz} shows the variation of $\Delta a_e$ with respect to the mass of the gauge boson $m_{Z_{e \tau}}$. The gridlines are $3 \sigma$ allowed values of $\Delta a_e$ in the unit of $10^{-13}$. From this figure, it can be seen that the allowed range of mass of the new gauge boson that satisfies the electron's anomalous magnetic moment is from 1 MeV to 150 MeV. 

For model B, similar relations are used to calculate electron $(g-2)$ with replacement of $Z_{e \tau} $ by $Z_{ e\mu}$; the corresponding Feynman diagram is shown in the left panel of Figure \ref{feyn} with $l^-=e^-$. The plot of $\Delta a_e$ with gauge boson mass ($m_{Z_{e \mu}}$) can be obtained for model B, which is found to have similar nature as Figure \ref{dae-mz}.

\subsection{Muon $(g-2)$}
\label{sub:mug-2}
 The  recent result from Fermilab's E989 experiment provides the most precise measurement of $a_\mu$ \cite{Muong-2:2023cdq},
  achieved through the combination of Run-1, Run-2 and Run-3 data as
  \begin{equation}
    a^{\rm FNAL}_\mu  = 116\hspace*{0.1 true cm}592\hspace*{0.1 true cm} 055 \hspace*{0.1 true cm}(24) \times 10^{-11}.  
  \end{equation}
Combining this result with those from the BNL  E$821$ experiment \cite{Bennett:2006fi},
provides the comprehensive experimental average
\begin{equation}
    a^{\rm exp}_\mu  = 116\hspace*{0.1 true cm}592\hspace*{0.1 true cm} 059 \hspace*{0.1 true cm}(22) \times 10^{-11}.  
  \end{equation}
Comparing this result with the SM prediction
\cite{Aoyama:2020ynm}
\begin{equation}
    a^{\rm SM}_\mu  = 116\hspace*{0.1 true cm}591\hspace*{0.1 true cm} 810 \hspace*{0.1 true cm}(43) \times 10^{-11},  
  \end{equation}
  yields a discrepancy of 5.1$\sigma$ \cite{Bennett:2006fi}, 
  \begin{equation}
     \Delta a_\mu = a^{\rm exp}_\mu - a^{\rm SM}_\mu = 2.49\hspace*{0.1 true cm}(0.48) \times 10^{-9}. 
  \end{equation}
  \iffalse
\begin{equation}
\Delta a^{\rm BNL}_\mu = a^{\rm exp}_\mu - a^{\rm SM}_\mu = (2.61\pm 7.9) \times 10^{-9}.
\end{equation}
\fi
These discrepancies from two independent experiments show clear failure of the SM, demonstrating the need to move beyond the standard model. Numerous models \cite{Altmannshofer:2016brv,Majumdar:2020xws,Mondal:2021vou,Hammad:2021mpl,Jegerlehner:2009ry,Moore:1984eg,Bauer:2018onh} have been proposed in the past to explain $(g-2)_\mu $. In the present work, model B can explain this discrepancy. The relevant interaction Lagrangian  is provided as follows \cite{Moore:1984eg}:
\begin{equation}
\mathcal{L}= -g_{e \mu}  \overline{\mu} \gamma^\mu \mu Z_{e \mu} - y_\Delta  \overline{\tau^c_L}  \mu_L \Delta^{++} + {\rm h.c.},
\end{equation}
where $g_{e \mu}$ and $Z_{e \mu} $ are the gauge parameters associated with $U(1)_{L_e-L_\mu} $ symmetry and $y_\Delta $ is the Yukawa coupling for the muon interaction with $\tau$ through the scalar  triplet $\Delta$, which breaks lepton number by two units, i.e., $\Delta L=2$. The possible Feynman diagrams which can provide additional contributions are shown in Figure \ref{feyn} with $l^-=\mu^-$. The expression for $a^{Z_{e\mu}}_\mu $  i.e., the gauge mediated contribution, can be deduced as follows \cite{Moore:1984eg}:
\begin{equation}
a^{Z_{e \mu}} _\mu =  \frac{g^2_{e \mu}}{4 \pi^2 } \int_0^1 \frac{x^2 (1-x)}{x^2 + \frac{m^2_{Z_{e \mu}}}{m^2_\mu }(1-x)}dx\;,
\label{eqn:amu_Zemu}
\end{equation}
where the mass of $Z_{e \mu} $ is on the MeV scale. Similarly, the contribution in the triplet portal is provided by \cite{Moore:1984eg}
\begin{equation}
a^{\Delta}_\mu= -\frac{m^2_{\mu} y^{2}_{\Delta}}{8 \pi^2} \int_0^1  \frac{\left( x^2-x^3+ \frac{m_{\tau}}{m_\mu} x^2 \right) + \left( x^2-x^3- \frac{m_{\tau}}{m_\mu} x^2\right) }{m^2_{\mu} x^2 + (m^2_{\tau}- m^2_\mu ) x + m^2_{\Delta} (1-x) } dx.
\label{eqn:del_amu}
\end{equation}

\begin{figure}[htb] 
\subfloat[]{\includegraphics[height=65mm,width=75mm]{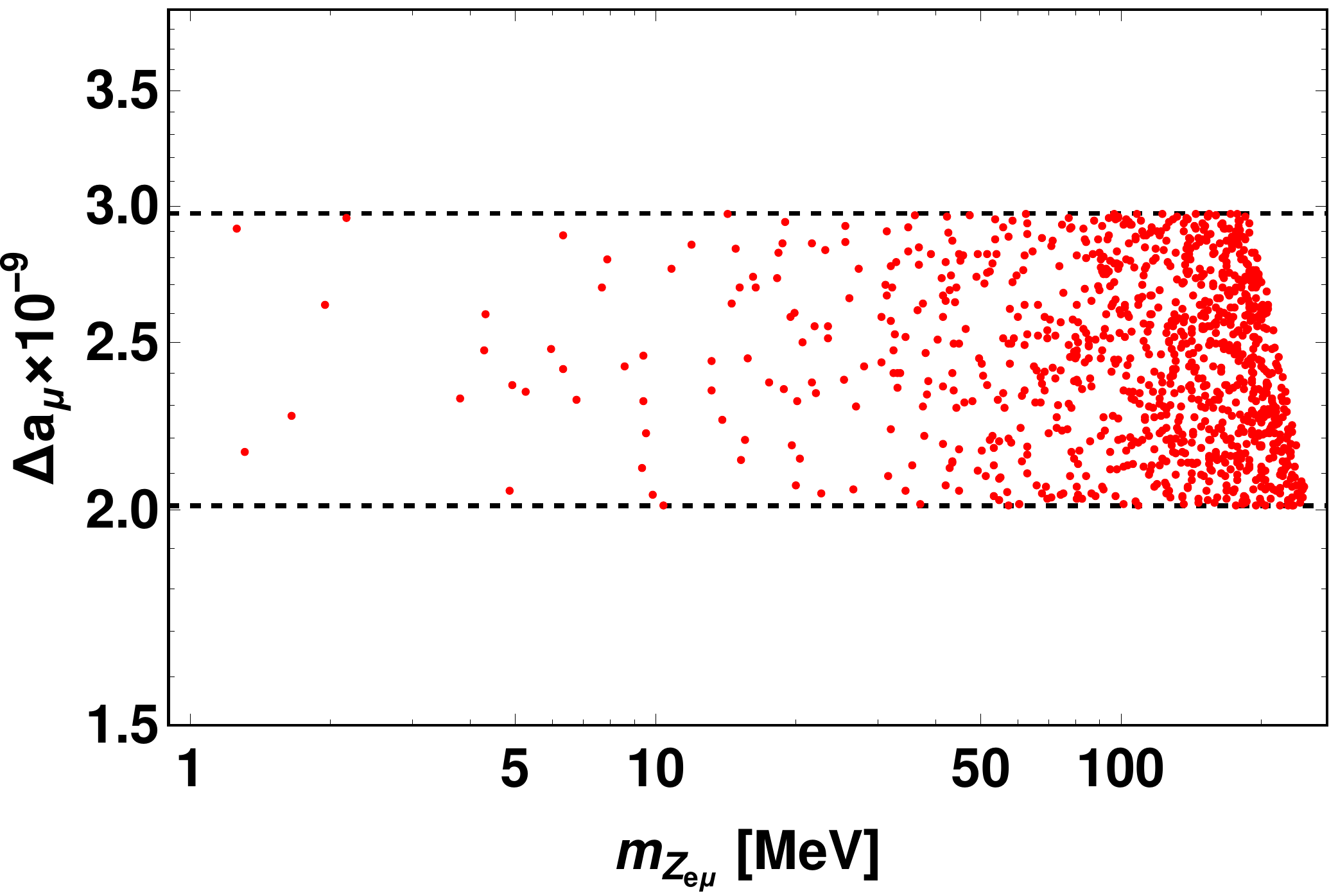}\label{fig-a}}
%\hspace{1cm}
\subfloat[]{\includegraphics[height=65mm,width=90mm]{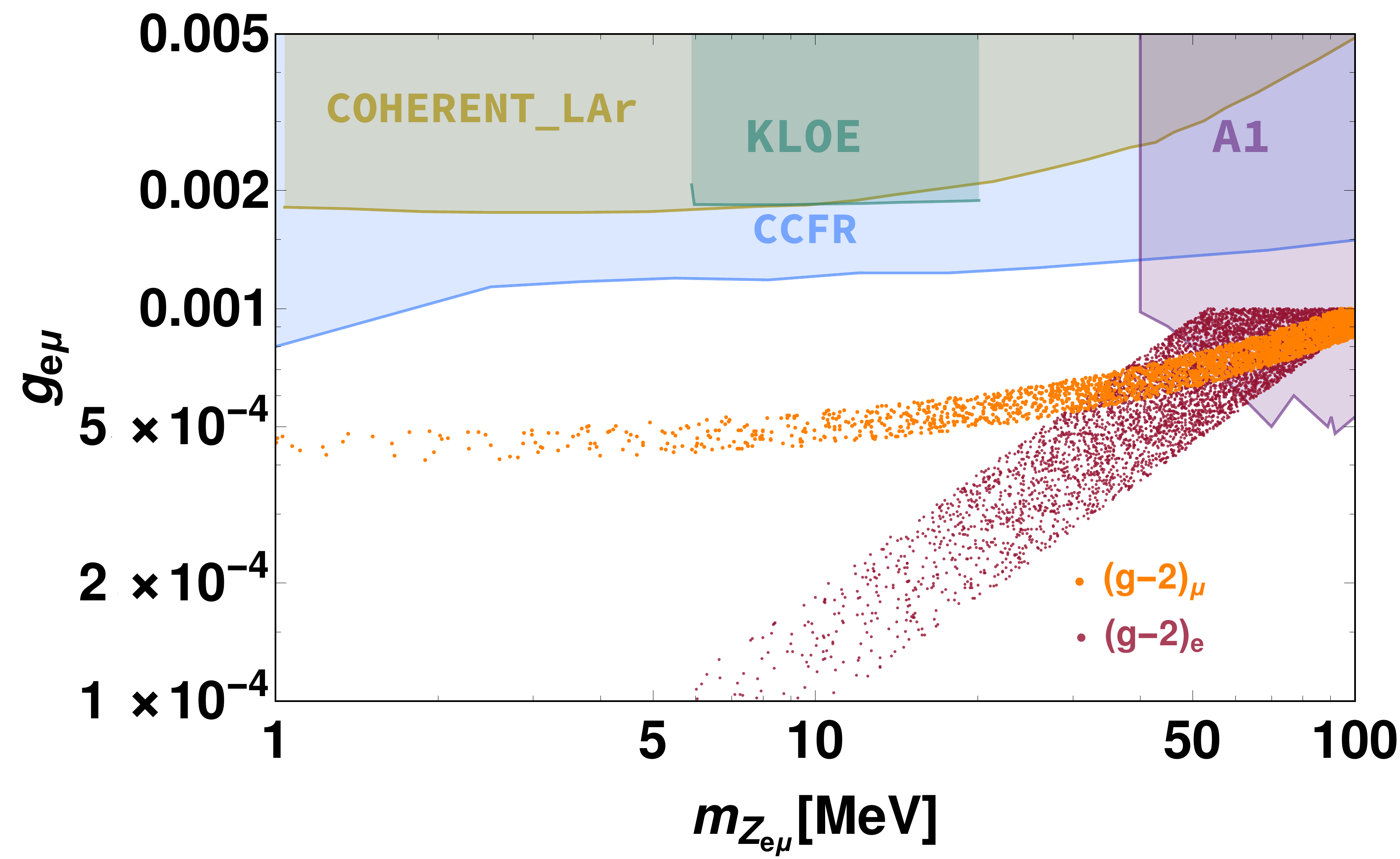}\label{fig-b}}
\caption{\textbf{(a)} Variation of $\Delta a_\mu$ value with respect to the mass of the gauge boson $(m_{Z_{e \mu}})$, \textbf{(b)} Variation of $g_{e \mu}$ with respect to gauge boson mass. In the plot, the brown and orange points are the outcomes from our model for electron and muon $(g-2)$, respectively while the blue, yellow, green and purple regions respectively represent  the experimental upper bounds of the CCFR, COHERENT$\_$LAr \cite{Borah:2021khc}, KLOE \cite{KLOE-2:2014qxg}, and A1\cite{PhysRevLett.106.251802,PhysRevLett.112.221802} experiments.}
\label{Fig:B-e-mu}
\end{figure}

As demonstrated above, neutrino phenomenology restricts the mass range of the scalar
triplet within $\sim$ ${\cal O}$(TeV) which is very large as compared to the muon and tau masses. Considering these assumptions, we have
\begin{equation}
 a^{\Delta}_\mu = -\frac{m^2_\mu y^{ 2}_\Delta }{12 \pi^2 m^2_{\Delta}}.
\label{eqn:del_amu_simplified}
\end{equation}
Equation (\ref{eqn:del_amu_simplified}) provides a feeble negative contribution towards the anomalous magnetic moment; however as $y_\Delta$ is very small we can safely neglect the term. Effectively, the extra contribution in muon $(g-2)$ is from the gauge boson only. Figure \ref{fig-a}  shows the variation of $\Delta a_\mu $ with respect to the mass of the gauge boson ($m_{Z_{e \mu}}$), while Figure \ref{fig-b} portrays the allowed region in the plane $g_{e \mu}-m_{Z_{e \mu}}$, consistent with Fermilab's  muon $(g-2)$ measurement \cite{Abi:2021gix} as well as the CCFR, neutrino trident bound \cite{conrad1998precision,Mishra:1991bv,Altmannshofer:2014pba}, COHERENT$\_$LAr bound \cite{Borah:2021khc}, KLOE \cite{KLOE-2:2014qxg} and A1\cite{PhysRevLett.106.251802,PhysRevLett.112.221802} bound.
\\
Summarizing the $(g-2)$ section, it is apparent that of the two proposed models
model A can explain electron $(g-2)$, whereas model B provides acceptable outcomes for both electron and muon $(g-2)$ discrepancies. Figure \ref{fig-b} provides a stringent bound on the parameter space of gauge coupling and the mass of the gauge boson $ Z_{e \mu}$ that are consistent with the current experimental bounds. In the above-mentioned mass range, the gauge boson can explain both electron and muon $(g-2)$ anomalies with great accuracy. Unlike $Z_{ e \mu}$, there is no such specific constraint on the mass of gauge boson $Z_{e \tau}$; however from  \cite{Bodas:2021fsy}, we can say that $m_{Z_{e \tau}}$ is in the $\rm{MeV}$ scale in order to explain $(g-2)_e$ and $(g-2)_\mu$ simultaneously.

\section{Concluding remarks}
\label{sec:con}
{The motivation behind the present work comes from the interest in exploring neutrino
phenomenology in $U(1)$ models. Therefore, we have investigated $U(1)_{L_e -L_\tau}$ (model A) as well as $U(1)_{L_e - L_\mu}$ (model B), which contain heavy neutral fermions and a scalar triplet to
realize a type-(I+II) hybrid seesaw. The imposition of $U(1)_{L_e - L_{\tau(\mu)}}$ gauge symmetry places limitations on lepton flavor structure, helping in the study of neutrino phenomenology. 
The general form of the neutrino mass matrix $m_\nu$ provides correct ranges of neutrino oscillation parameters.
 From the neutrino phenomenology section,it can be seen that both the models provide the full $3 \sigma$ range of the NuFit v5.2 data for the oscillation parameters $\theta_{13}, \theta_{12}$ and $\theta_{23}$. 
 Interestingly, Model A imposes a strong constraint on the value of $\delta_{\rm CP}$, favoring CP-conserving values primarily around $0^\circ$ and $360^\circ$; in contrast, model B allows for a much broader range of $\delta_{\rm CP}$, spanning from $20^\circ$ to $350^\circ$.

Our plots place bounds on the sum of neutrino mass as follows: $0.043 ~(0.058)~\rm{eV} \lesssim\sum_i m_i \lesssim 0.114~(0.107) ~\rm eV$ for model A (model B). 
 We show the capacity of the future long-baseline neutrino experiments DUNE, P2SO, T2HK, and T2HKK to test our proposed models. 
 The results indicate that DUNE and T2HKK can probe only model B within their $5\sigma$ parameter space in the $\theta_{23}^{\rm true}-\delta_{\rm CP}^{\rm true}$ plane; in contrast, P2SO and T2HK have the potential to probe both models within their $5\sigma$ parameter space if the NuFit v5.2 oscillation parameters remain as the true values in the future. Apart from this, our models successfully explain muon and electron $(g-2)$; models A and B both demonstrate electron $(g-2)$, while model B showcases admissible results for muon $(g-2)$ that are within the bounds set by the CCFR, KLOE, A1 and COHERENT$\_$LAr} experiments.

 Concerning to relevance of our models and potential future insights, if heavy fermions
can be made stable with some symmetry, they can contribute to the dark matter relic density
present in the Universe. Through coupling with the new $U(1)$ gauge boson and scalars,
they can annihilate through weak interaction and contribute to the DM budget as Weakly Interacting Massive Particles (WIMPs), and can also provide a direct signal in detection
experiments. Dual portal annihilation channels can provide new region of the DM mass
spectrum as well as a constrained parameter space to probe in future experiments in search
of DM. Moreover, correlative studies investigating neutrino oscillation phenomena could
be carried out as well.

\section*{Acknowledgements}
 PP and PM would like to acknowledge Prime Minister's Research Fellowship for its financial help. MKB acknowledge support from the NSRF via the Program Management Unit for Human Resources \& Institutional Development, Research, and Innovation [grant no. B13F660066]. MKB also acknowledge the National Science and Technology Development Agency, National e-Science Infrastructure Consortium, Chulalongkorn University, and the Chulalongkorn Academic Advancement into Its 2nd Century Project (Thailand) for providing computing infrastructure that has contributed to the results reported within this paper. SS and RM would like to acknowledge University of Hyderabad IoE project grant no. RC1-20-012. 

\bibliographystyle{my-JHEP}
\bibliography{e-j}

\end{document}